\documentclass[lettersize,journal]{IEEEtran}
\usepackage{amsmath, amssymb, amsfonts}

\usepackage{algorithm}
\usepackage{algpseudocode} 

\usepackage{graphicx}
\usepackage{epstopdf} 
\usepackage{booktabs} 
\usepackage{array}
\usepackage{multirow}
\usepackage{makecell}
\usepackage{diagbox}
\usepackage{float} 
\usepackage{lscape} 
\usepackage[caption=false,font=normalsize,labelfont=sf,textfont=sf]{subfig} 

\usepackage{xcolor}
\usepackage{textcomp}
\usepackage{stfloats}
\usepackage{url}
\usepackage{verbatim}
\usepackage{cite}
\usepackage{ragged2e}
\usepackage{cuted} 

\usepackage[colorlinks, linkcolor=red, anchorcolor=blue, citecolor=green]{hyperref}

\usepackage{amsthm}

\begin{document}

\title{Safe Multi-Agent Deep Reinforcement Learning for Privacy-Aware Edge-Device Collaborative DNN Inference}

\author{Hong~Wang, Xuwei~Fan, Zhipeng~Cheng, Yachao~Yuan, Minghui~Min, \textit{Senior Member, IEEE},
        Minghui~Liwang, \textit{Senior Member, IEEE}, Xiaoyu~Xia, \textit{Senior Member, IEEE}
\thanks{Hong Wang (wh\_5233@163.com), Zhipeng Cheng (chengzp\_x@163.com) and Yachao Yuan (chao910904@suda.edu.cn) are with School of Future Science and Engineering, Soochow University, Suzhou 215006, China. Xuwei Fan (xwfan@fafu.edu.cn) is with College of Computer and Information Sciences, Fujian Agriculture and Forestry University, Fuzhou 350002, China. Minghui Min (minmh@cumt.edu.cn) is with the School of Information and Control Engineering, China University of Mining and Technology, Xuzhou 221116, China. Minghui Liwang (minghuilwang@tongji.edu.cn) is with the Department of Control Science and Engineering, Tongji University, Shanghai 201804, China. Xiaoyu Xia (xiaoyu.xia@rmit.edu.au) is with the School of Computing Technologies, RMIT University, Melbourne, Victoria, Australia. (Hong~Wang and Xuwei~Fan contribute equally to this work. \textit{Corresponding author: Zhipeng Cheng})}}

\markboth{Journal of \LaTeX\ Class Files,~Vol.~14, No.~8, August~2021}%
{Shell \MakeLowercase{\textit{et al.}}: A Sample Article Using IEEEtran.cls for IEEE Journals}


\maketitle

\begin{abstract}
As Deep Neural Network (DNN) inference becomes increasingly prevalent on edge and mobile platforms, critical challenges emerge in privacy protection, resource constraints, and dynamic model deployment. This paper proposes a privacy-aware collaborative inference framework, in which adaptive model partitioning is performed across edge devices and servers. To jointly optimize inference delay, energy consumption, and privacy cost under dynamic service demands and resource constraints, we formulate the joint problem as a Constrained Markov Decision Process (CMDP) that integrates model deployment, user-server association, model partitioning, and resource allocation. We propose a Hierarchical Constrained Multi-Agent Proximal Policy Optimization with Lagrangian relaxation (HC-MAPPO-L) algorithm, a safe reinforcement learning-based framework that enhances Multi-Agent Proximal Policy Optimization (MAPPO) with adaptive Lagrangian dual updates to enforce long-term delay constraints. To ensure tractability while maintaining coordination, we decompose the CMDP into three hierarchically structured policy layers: an auto-regressive based model deployment policy, a Lagrangian-enhanced user association and model partitioning policy, and an attention-based resource allocation policy. Extensive experimental results demonstrate that HC-MAPPO-L consistently satisfies stringent delay constraints while achieving a superior balance among energy consumption and privacy cost, outperforming representative baseline algorithms across varying problem scales and resource configurations.
\end{abstract}


\section{Introduction}\label{sec:intro}
Large-scale Deep Neural Networks (DNNs) are increasingly deployed in delay-sensitive applications such as autonomous driving, smart healthcare, and real-time video analytics. These models create substantial computational demands that overwhelm resource-constrained end devices~\cite{FanICC, Zhang2023EERJSAC}. To bridge this resource gap without incurring the high latency and privacy risks of full cloud offloading, edge-device collaborative inference via DNN model partitioning has emerged as a key paradigm~\cite{Li2022PrivacyMEC}. This approach, supported by Mobile Edge Computing (MEC) infrastructure, strategically splits a DNN by executing initial layers on the device and offloading subsequent layers to edge servers, thereby reducing delay and bandwidth usage compared to full cloud execution~\cite{9296560, Zhang2024GenerativeAIJSAC}. Pioneering systems like \emph{Neurosurgeon}~\cite{Kang2017Neurosurgeon} established the viability of dynamic model partitioning, demonstrating that partition points can adapt to network and device conditions. The effectiveness of such collaborative inference has been further validated by subsequent designs~\cite{Zhang2021DeepSlicing}, with recent surveys comprehensively documenting its expansion~\cite{Ren2023}.

However, most existing studies on collaborative inference primarily focus on optimizing multi-dimensional Quality of Service (QoS) objectives, such as inference delay, energy consumption, and inference accuracy, often underestimating a critical vulnerability: privacy leakage. Transmitting intermediate features to edge servers does not eliminate risks; these features remain susceptible to inversion attacks~\cite{Erdogan2022UnSplit}. As earlier-layer activations retain rich spatial and semantic information about the raw input, they can be exploited to reconstruct sensitive data, especially when augmented with side information~\cite{mao2022secure}. Therefore, privacy must be treated not as an afterthought, but as an intrinsic and coequal objective in the model partitioning and scheduling problem.

Recent efforts incorporate privacy-preserving techniques such as encryption, differential privacy, and fixed partitioning~\cite{Yao2024SecoInfer}. Nevertheless, practical deployments of collaborative inference face intertwined challenges arising from the inherent trade-off among privacy, latency, and resource constraints: improving on-device privacy increases computational latency, whereas aggressive offloading reduces delay but exposes sensitive features. First, there is a notable absence of structure-adaptive and privacy-aware joint optimization. Most works often treat privacy as a secondary constraint rather than an integral objective~\cite{10464411}. Crucially, the layer-wise privacy sensitivity of DNNs and its intricate coupling with service caching, user association, and resource allocation decisions are seldom modeled, preventing a systematic trade-off between privacy and efficiency under dynamic conditions.

A second challenge lies in achieving efficient resource allocation under multi-dimensional QoS constraints. Many existing approaches either optimize for a single metric (e.g., delay) or rely on strict per-task guarantees, ill-suited to applications demanding coupled guarantees across latency, energy, and privacy~\cite{Zhang2024InteractiveAINetwork,8269175}. The former sacrifices other critical dimensions, while the latter leads to resource underutilization due to worst-case provisioning. A more robust strategy is to model latency as a long-term average constraint, which improves resource efficiency by allowing occasional, tolerable violations while ensuring overall service quality.

Finally, ensuring constraint satisfaction in Deep Reinforcement Learning (DRL)-based optimization remains a significant hurdle. While DRL is well-suited for dynamic edge environments, most algorithms handle constraints by incorporating violation penalties into the reward function. This reward-penalty coupling often causes training instability or even prevents convergence under stringent constraints, as agents can be discouraged by persistently low rewards, hindering effective policy exploration. In contrast, safe DRL explicitly integrates safety constraints into the policy optimization process~\cite{10444044}, enabling the agent to maximize long-term performance while ensuring constraint satisfaction and maintaining policy feasibility and stability. This makes safe DRL particularly necessary for privacy-aware collaborative inference, where delay constraints must be satisfied under the coupled privacy--resource trade-off. However, effectively integrating safe DRL into large-scale multi-agent collaborative inference systems remains a formidable challenge~\cite{zhu2024safe}.

To address these challenges, we formulate the problem as a Constrained Markov Decision Process (CMDP) and propose the Hierarchical Constrained Multi-Agent Proximal Policy Optimization with Lagrangian relaxation (HC-MAPPO-L), a novel safe reinforcement learning-based hierarchical framework. This framework naturally aligns with the edge-device collaborative inference system by decomposing the problem into complementary decision layers: slow-timescale model deployment and fast-timescale operational decisions. The framework is further strengthened by an adaptive Lagrangian mechanism that rigorously enforces the long-term delay constraint while ensuring stable training. The main contributions of this paper are summarized as follows:
\begin{enumerate}
    \item We establish a comprehensive optimization framework for privacy-aware edge-device collaborative inference, which is formally cast as a novel CMDP. This formulation jointly addresses model deployment, user-server association, privacy-aware model partitioning, and multi-dimensional resource allocation, incorporating explicit quantitative models for energy consumption and privacy leakage under a long-term average delay constraint.
    \item We propose a HC-MAPPO-L algorithm that integrates adaptive Lagrangian dual updates to enforce long-term delay guarantees. The algorithm features a multi-timescale architecture with specialized policies, including an auto-regressive based model deployment policy, a Lagrangian-enhanced joint association-partitioning policy, and an attention-based allocation policy, which enables efficient and scalable decision-making.
    \item We conduct extensive experiments that validate the robustness and superiority of the proposed algorithm. Simulation results demonstrate that HC-MAPPO-L consistently satisfies stringent delay constraints while achieving a superior trade-off among system cost, energy consumption, and privacy preservation, outperforming state-of-the-art approaches across a wide range of system scales and sensitivity tests.
\end{enumerate}

The remainder of this paper is organized as follows. Section II reviews related work. Section III describes the system model. Section IV details the proposed algorithm. Section V presents evaluation results, and Section VI concludes.

\section{Related Work}\label{sec:related_work}

This section reviews prior research closely related to our work, focusing on three key areas: (i) model partitioning and deployment for collaborative inference, (ii) privacy preservation in collaborative inference, and (iii) safe Multi-Agent Reinforcement Learning (MARL) with constraint handling.

\subsection{Model Partitioning and Deployment for Collaborative Inference}
Partitioning DNNs between resource-constrained devices and servers has proven effective for meeting latency and energy budgets. Pioneering this line of work,~\cite{Kang2017Neurosurgeon} introduced layer-wise profiling for end-cloud model partitioning, demonstrating that optimal partition points depend on runtime bandwidth and device capabilities. This approach was later extended to end-edge-cloud hierarchies by~\cite{7979979}, which incorporated early exits to show that hierarchical collaboration can significantly reduce end-to-end delay without compromising accuracy. Moreover, efficient service caching and model deployment are critical complements to partitioning strategies. As identified in~\cite{Mao2017MECSurveyComm}, these techniques are essential for reducing latency and conserving network bandwidth. Recent work explores two-timescale optimization frameworks to address the coupling between long-term deployment and short-term resource allocation.~\cite{10609797} proposed a DRL-based approach with long-timescale service caching policies to minimize switching costs and short-timescale resource management to maximize QoS. Similarly,~\cite{Zheng2025MultiDNN} jointly optimized model placement, request scheduling, and multi-dimensional resource allocation for multi-DNN parallel inference in end-edge-cloud networks.

Based on these foundations, subsequent research has improved runtime adaptability through enhanced profiling techniques, dynamic scheduling mechanisms, and feature compression methods. Recent advances in adaptive model partitioning combine fine-grained partitioning with strategies like model pruning~\cite{Li2025AdaptiveMEC,Niazmand2025TCCN} and inference parallelism~\cite{Li2023DelayAware} to maximize throughput under strict delay constraints. However, the joint optimization of multiple QoS metrics, including latency, energy, and reliability, remains challenging. Current state-of-the-art approaches typically formulate this challenge as a utility maximization problem under multi-dimensional resource constraints, exemplified by parallel offloading with load balancing~\cite{Zhang2022CoopMEC,Ye2018TVT,Ye2018VTMag} and integrated mode selection with resource allocation in vehicular networks~\cite{Zhang2024PartialOffload,Ye2021OJVT}. Despite these advances, most existing systems focus on optimizing isolated components, such as partitioning under fixed deployment configurations or resource allocation for predetermined placements. This fragmented methodology overlooks the critical interdependence among cache placement strategies, privacy-aware user association, and dynamic resource allocation, while also failing to provide explicit long-term service guarantees. Our hierarchical framework systematically addresses these limitations by coordinating these decisions across multiple timescales.

\subsection{Privacy Preservation in Collaborative Inference}
Privacy preservation in MEC generally encompasses location privacy, usage pattern obfuscation, and secure data transmission. In the context of model partitioning, although raw data remains on the device, the intermediate activations transmitted to the edge can still leak sensitive information. For example,~\cite{Dosovitskiy_2016_CVPR} revealed that internal features preserve spatial or semantic patterns that allow high-fidelity input reconstruction, while honest-but-curious servers may combine model inversion with stealing attacks~\cite{Erdogan2022UnSplit}. These findings collectively suggest that shallow partitions are inherently more privacy-sensitive than deeper ones, motivating the need for adaptive feature compression.

Quantifying privacy leakage remains challenging. The Structural Similarity index (SSIM) offers a perceptual, differentiable proxy for reconstruction fidelity, serving as a practical leakage metric~\cite{Wang2004SSIM}. To mitigate risks, privacy-aware adaptive partitioning~\cite{Jiang2023PrivacyAware} dynamically adjusts partition points to balance privacy, communication, and computation costs. DataMix~\cite{Liu2020DataMix} employs mixup-inspired transformations to protect offloaded data, demonstrating that privacy mechanisms can coexist with efficient delegation. Other defense strategies include feature perturbation and privacy-preserving activations~\cite{mao2022secure}, which further balance privacy and utility. However, these approaches typically consider single device-edge pairs. In contrast, our formulation embeds SSIM-based leakage cost into the learning objective, enabling on-the-fly trade-offs among privacy, latency, and energy under multi-user resource constraints.

\subsection{Safe MARL with Constraint Handling}

MARL under the Centralized Training with Decentralized Execution (CTDE) paradigm has demonstrated strong scalability in cooperative edge computing tasks~\cite{9462346}. In this framework, agents make decisions based on local observations during execution while critics utilize global information for training, enabling effective coordination through value factorization and actor-critic architectures. However, conventional MARL methods primarily maximize expected returns without adequately addressing hard service constraints, limiting their practical applicability. Recent advancements have sought to enhance MARL's capabilities in edge environments: attention mechanisms enable handling of variable-dimensional observations and actions~\cite{Li2025AttentionMARL}, while hierarchical frameworks like federated learning with intermediate aggregators effectively manage system heterogeneity~\cite{app12020670}.

When considering safety requirements, CMDP provides the foundational framework~\cite{Altman1999CMDP}, with established solution approaches including primal-dual methods, constrained policy optimization (CPO)~\cite{Achiam2017CPO}, and Lyapunov-based optimization~\cite{chow2019lyapunovbasedsafepolicyoptimization}. These techniques have been successfully applied to MEC scenarios such as reliable task offloading~\cite{8269175} and parallel offloading with load balancing~\cite{Zhang2022CoopMEC}. Extending constraint handling to multi-agent settings introduces additional challenges from partial observability, action coupling, and non-stationarity, where CTDE provides coordination benefits but does not guarantee constraint feasibility. To address these limitations, we develop a constrained MARL framework that incorporates Lagrangian relaxation within a hierarchical architecture. Unlike barrier, projection, or trust-region methods that incur conservatism or high computational overhead, the proposed Lagrangian-based approach enforces long-term delay constraints through lightweight dual updates, while preserving the scalability required in dynamic edge environments and aligning MARL coordination with the requirements of privacy-aware collaborative inference.

\section{System Model}\label{sec:system}
\subsection{Network Model}
\begin{figure}[t]
\centering
\includegraphics[width=3in]{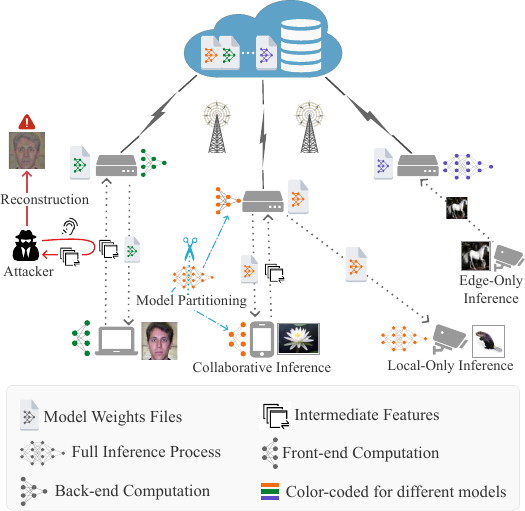}
\caption{The system model of edge-device collaborative inference.}
\label{fig:edge_computing_scenario}
\end{figure}

As illustrated in Fig.~\ref{fig:edge_computing_scenario}, we consider a hierarchical edge computing system composed of a central cloud, $J$ distributed edge servers in the set $\mathcal{J} = \{1,2,\ldots,j,\ldots,J\}$, and $K$ heterogeneous user devices in $\mathcal{K} = \{1,2,\ldots,k,\ldots,K\}$. The cloud acts as a model repository, storing $I$ distinct pre-trained DNN models (denoted by $\mathcal{I} = \{1,2,\ldots,i,\ldots,I\}$), which are cached by edge servers for rapid local access. When a user device generates an inference task, the system employs a collaborative inference paradigm where the DNN model execution is dynamically partitioned: the device downloads front-end parameters from an associated server, performs local computation, and then offloads intermediate features for the server to complete the back-end inference. The choice of this partition point jointly determines the overall delay, energy consumption, and privacy cost.

The system operates over discrete time slots $\mathcal{T} = \{1,2,\ldots,t,\ldots,T\}$. To maintain operational stability, model deployment decisions occur at a slower timescale. Specifically, each server $j$ updates every ${\scriptstyle\Delta}T$ time slots. The model deployment decision is denoted by a binary variable $x_{i,j}(t)$, where $x_{i,j}(t)=1$ indicates model $i$ is deployed on server $j$, otherwise $x_{i,j}(t)=0$. The complete deployment configuration $X(t) = \{x_{i,j}(t) \mid i \in \mathcal{I}, j \in \mathcal{J}\}$ is held constant throughout each ${\scriptstyle\Delta}T$-slot interval.

Within each time slot $t$, users generate service requests $Req_k(t) = (i, D_k^{\mathrm{in}}(t))$ specifying the requested model $i$ and input batch size $D_k^{\mathrm{in}}(t)$ (i.e., number of data samples for inference). For collaborative inference, user $k$ first selects an edge server through binary variable $y_{j,k}(t)$, where $y_{j,k}(t) = 1$ indicates the connection to server $j$, otherwise $y_{j,k}(t) = 0$. In addition, let $\sum_{j \in \mathcal{J}} y_{j,k}(t) = 1$ to ensure that each user can associate with exactly one server at each time slot. The set of all such association decisions is denoted by $Y(t) = \{y_{j,k}(t) \mid j \in \mathcal{J}, k \in \mathcal{K}\}$. Based on these decisions, the associated servers then allocate corresponding computing and bandwidth resources.

\subsection{DNN Partitioning Model}\label{sec:splitting_model}

The selection of DNN partitioning points fundamentally governs the trade-off between computational workload and privacy protection. Following the layer-wise partitioning paradigm~\cite{Kang2017Neurosurgeon}, our framework strategically divides DNN execution between devices and edge servers. Deeper partitions enhance privacy by utilizing more abstract features, albeit at the cost of increased on-device energy consumption. This subsection formalizes the partitioning strategy and quantifies its system-wide implications.

\begin{figure}[t]
\centering
\includegraphics[width=3.0in]{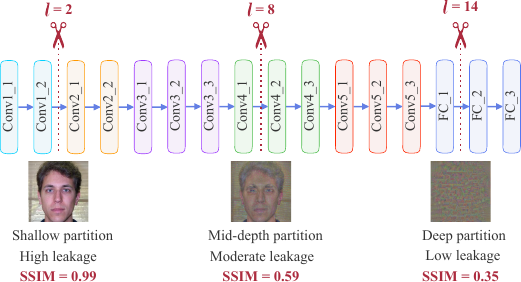}
\caption{Privacy leakage decreases with partition depth in VGG16, showing SSIM scores of 0.99 ($l = 2$), 0.59 ($l = 8$), and 0.35 ($l = 14$).}
\label{fig:vgg16_splitting}
\end{figure}

For each DNN model $i$, we define the partitioning decision as $z_{i,k}(t) \in \{0, 1, \ldots, L_i\}$, where $L_i$ denotes the number of partitionable layers.\footnote{Our partitioning considers only computational layers (e.g., convolutional, fully-connected) that dominate inference overhead. Non-computational layers (e.g., activation, pooling) are treated as atomic units with their preceding computational layers to maintain functional coherence and simplify the decision space.} A decision $z_{i,k}(t) = l$ indicates partitioning after layer $l$, with the first $l$ layers executing on the user device and the remaining $L_i-l$ layers on the edge server. The boundary cases $z_{i,k}(t) = 0$ and $z_{i,k}(t) = L_i$ represent full edge and full device execution, respectively. All user decisions form the partitioning set $Z(t) = \{z_{i,k}(t) \mid i \in \mathcal{I}, k \in \mathcal{K}\}$. Fig.~\ref{fig:vgg16_splitting} demonstrates this partitioning scheme using VGG16, quantitatively showing how deeper partitions reduce SSIM scores and thereby strengthen privacy protection.

The partitioning decision $z_{i,k}(t)$ directly governs five critical performance parameters in collaborative inference: parameter download volume $D_k^{\mathrm{U}}(t)$ for device-side layers; on-device computational workload $\Omega_k^{\mathrm{U}}(t)$; intermediate feature upload volume $D_k^{\mathrm{up}}(t)$; edge-side computational workload $\Omega_k^{\mathrm{E}}(t)$; and privacy leakage risk $Lea_k(t)$.

To operationalize these parameters, we conduct layer-wise profiling of each DNN model using a representative unit input. For layer $l$ in model $i$, we define its computational load as $\omega_{l,i}$ (measured in Floating Point Operations, FLOPs), parameter volume as $S_{l,i}$, and output feature volume as $D_{l,i}^{\mathrm{out}}$. For convolutional layers, the feature map data volume is computed as:
\begin{equation}\label{eq:conv_data}
D_{l,i}^{\mathrm{out}} = D_i^{\mathrm{elem}} \cdot H_{l,i}^{\mathrm{out}} \cdot W_{l,i}^{\mathrm{out}} \cdot C_{l,i}^{\mathrm{out}}
\end{equation}
where $D_i^{\mathrm{elem}} = 4$ bytes denotes single-precision element size, and $H_{l,i}^{\mathrm{out}}, W_{l,i}^{\mathrm{out}}, C_{l,i}^{\mathrm{out}}$ represent output feature dimensions. For fully connected layers:
\begin{equation}\label{eq:dense_data}
D_{l,i}^{\mathrm{out}} = D_i^{\mathrm{elem}} \cdot V_{l,i}^{\mathrm{out}}
\end{equation}
where $V_{l,i}^{\mathrm{out}}$ is the output vector length.

Based on these layer-wise metrics, the five key system parameters at partition point $z_{i,k}(t)$ are formally defined. The parameter download volume is computed as $D_k^{\mathrm{U}}(t) = \sum_{m=1}^{z_{i,k}(t)} S_{m,i}$, representing the cumulative volume of device-side layers. The local computational workload equals $\Omega_k^{\mathrm{U}}(t) = \sum_{m=1}^{z_{i,k}(t)} \omega_{m,i}$, quantifying the total FLOPs for these layers. The remaining edge-side computation is given by $\Omega_k^{\mathrm{E}}(t) = \sum_{m=z_{i,k}(t)+1}^{L_i} \omega_{m,i}$, constituting the server's workload. Finally, the upload data volume corresponds to $D_k^{\mathrm{up}}(t) = D_{z_{i,k}(t),i}^{\mathrm{out}}$, which is the output volume of the final locally-processed layer.

For privacy quantification, we adopt the SSIM in~\cite{Wang2004SSIM} as a practical perceptual metric for feature reconstructability. The privacy-leakage score is defined as:
\begin{equation}\label{eq:ssim_privacy_leakage}
Lea_k(t) = \mathrm{SSIM}(I_o, I_r|z_{i,k}(t))
\end{equation}
where $I_o$ denotes the original input, $I_r|z_{i,k}(t)$ represents the image reconstructed from intermediate features at partition depth $z_{i,k}(t)$, and $\mathrm{SSIM}(\cdot) \in [0,1]$. This formulation captures the empirical observation that shallow layers preserve low-level information, whereas deeper layers rely on more abstract representations. We employ SSIM instead of alternatives such as mutual information or attack success rates, as it provides a computationally efficient and attacker-agnostic proxy for feature reconstructability. Moreover, the layer-wise leakage profile in Appendix~B shows that privacy gains diminish beyond moderate depths, motivating the selection of partition points that balance privacy protection with computation and transmission costs. We define the boundary cases as follows: when $z_{i,k}(t) = 0$, indicating full edge execution with maximum privacy risk, $Lea_k(t) = 1$; when $z_{i,k}(t) = L_i$, representing full device execution with complete privacy protection, $Lea_k(t) = 0$.

\subsection{Inference Delay Model}
We assume quasi-static channels within each time slot, for highly dynamic environments (e.g., vehicular networks), the framework can be extended by incorporating mobility prediction or handover-aware association to maintain robustness. The end-to-end inference delay encompasses four sequential components: model download delay $\tau_k^{\mathrm{E2U}}$, local computation delay $\tau_k^{\mathrm{U}}$, feature upload delay $\tau_k^{\mathrm{U2E}}$, and edge computation delay $\tau_k^{\mathrm{E}}$.

To begin local execution of the first $z_{i,k}(t)$ layers, the device downloads corresponding parameters from edge server $j$ with delay $\tau_k^{\mathrm{E2U}}(t) = D_k^{\mathrm{U}}(t)/R_{j,k}^{\mathrm{E2U}}(t)$, where the downlink rate follows Shannon's formula $R_{j,k}^{\mathrm{E2U}} = B_{j,k}\log_2(1+P_j g_{j,k}/\sigma^2)$. Here $B_{j,k}$ denotes the allocated bandwidth of edge server $j$ for user $k$, $P_j$ is the server transmission power, $g_{j,k}$ incorporates path loss and shadow fading~\cite{Erceg1999PathLoss}, and $\sigma^2$ represents the noise power.

Local computation then processes the initial layers with delay $\tau_k^{\mathrm{U}}(t) = D_k^{\mathrm{in}}(t) \cdot \Omega_k^{\mathrm{U}}(t)/f_k$, where $f_k$ is the device's computational capacity. Subsequently, intermediate features are uploaded with delay $\tau_k^{\mathrm{U2E}}(t) = D_k^{\mathrm{in}}(t) \cdot D_k^{\mathrm{up}}(t)/R_{j,k}^{\mathrm{U2E}}(t)$, where the uplink rate $R_{j,k}^{\mathrm{U2E}} = B_{j,k}\log_2(1+P_k g_{j,k}/\sigma^2)$ depends on the user transmission power $P_k$.

The total delay for service success is composed of these components:
\begin{equation}
\tau_k^{\mathrm{succ}}(t) = \tau_k^{\mathrm{E2U}}(t) + \tau_k^{\mathrm{U}}(t) + \tau_k^{\mathrm{U2E}}(t) + \tau_k^{\mathrm{E}}(t)
\end{equation}

A service failure occurs if the model is not deployed (i.e., $x_{i,j}(t) = 0$ and $y_{j,k}(t) = 1$). To account for this, we define the total inference delay for user $k$ as:
\begin{equation}\label{eq:total_delay}
\tau_k(t) =
\begin{cases}
\tau_k^{\mathrm{succ}}(t), & \mathrm{if~} x_{i,j}(t) \cdot y_{j,k}(t) = 1 \\
\tau_{\mathrm{fail}}, & \mathrm{otherwise}
\end{cases}
\end{equation}
where $\tau_{\mathrm{fail}}$ is a prohibitively large constant representing service unavailability, typically set much larger than the delay constraint $\bar{\tau}$.

\subsection{Inference Energy Consumption Model}

We focus on energy consumption at user devices, as edge servers typically have stable power supplies. The energy expenditure comprises two main components: computation and transmission. The computational energy consumed by local inference is given by:
\begin{equation}
e_k^{\mathrm{U}}(t) = \epsilon \cdot D_k^{\mathrm{in}}(t) \cdot \Omega_k^{\mathrm{U}}(t)
\end{equation}
where $\epsilon$ denotes the energy coefficient in joules per FLOP~\cite{Horowitz2014Energy},~\cite{Tu2024EdgeEnergy}.

The transmission energy for uploading intermediate features is calculated as:
\begin{equation}
e_k^{\mathrm{U2E}}(t) = P_k \cdot \tau_k^{\mathrm{U2E}}(t)
\end{equation}
where $P_k$ is the user's transmission power and $\tau_k^{\mathrm{U2E}}(t)$ is the upload duration.

The total energy consumption for user $k$ during time slot $t$ combines these two components:
\begin{equation}\label{eq:user_energy}
e_k(t) = e_k^{\mathrm{U}}(t) + e_k^{\mathrm{U2E}}(t)
\end{equation}

To evaluate the system-wide energy performance, we employ the average per-user energy consumption:
\begin{equation}\label{eq:avg_energy}
\bar{e}(t) = \frac{1}{K}\sum_{k=1}^{K} e_k(t)
\end{equation}

\subsection{Privacy Cost Model}

As introduced in Section~\ref{sec:splitting_model}, the privacy leakage probability $Lea_k(t)$, measured by the Structural Similarity (SSIM) index, reflects the risk of input reconstruction at a chosen model partition point. To effectively incorporate this risk as a quantifiable metric within our optimization framework, we formulate a privacy cost model. This model accounts not only for the leakage probability itself, but also for the scale of processed data and the specific privacy preferences of individual users. Formally, the privacy cost for user $k$ at time slot $t$ is defined as:
\begin{equation}\label{eq:privacy_cost}
C_k^{\mathrm{priv}}(t) = (\alpha_1 + \alpha_2 \cdot Pre_k) \cdot Lea_k(t) \cdot D_k^{\mathrm{in}}(t) \cdot D_i^{\mathrm{raw}}
\end{equation}
Here, $Pre_k \in [0,1]$ denotes the privacy preference level of user $k$, and $D_i^{\mathrm{raw}}$ indicates the raw input data size per sample for model $i$ (e.g., the size of one image in bytes). The non-negative coefficients $\alpha_1$ and $\alpha_2$ control the baseline privacy cost and the sensitivity to user-specific privacy preferences, respectively.

To assess system-wide privacy performance, we evaluate the average per-user privacy cost across all users:
\begin{equation}\label{eq:avg_privacy}
\bar{C}^{\mathrm{priv}}(t) = \frac{1}{K}\sum_{k=1}^{K} C_k^{\mathrm{priv}}(t)
\end{equation}

\subsection{Problem Formulation}

Our objective is to minimize the long-term weighted sum of the average per-user energy consumption and privacy cost, under a constraint on the long-term average inference delay. Concretely, the instantaneous system cost at time slot $t$ is defined as:
\begin{equation}
\mathrm{Cost}(t) = \mu_1 \cdot \bar{C}^{\mathrm{priv}}(t) + \mu_2 \cdot \bar{e}(t)
\end{equation}
where $\mu_1$ and $\mu_2$ are non-negative weighting coefficients that balance the trade-off between privacy preservation and energy efficiency.

The overall optimization problem is thus formulated as minimizing the long-term average of this cost, subject to a long-term average delay constraint on the system-wide average per-user delay:
\allowdisplaybreaks
\begin{align}
\mathcal{P}: \quad &\min_{\substack{X(t),Y(t),Z(t),\\F(t),B(t)}} \lim_{T \rightarrow \infty} \frac{1}{T} \sum_{t=0}^{T-1} \mathbb{E}[\mathrm{Cost}(t)] \label{eq:optimization_problem} \\
&\mathrm{s.t.} \nonumber \\
&\quad x_{i,j}(t) \in \{0,1\}, \quad \forall i \in \mathcal{I}, \forall j \in \mathcal{J} \tag{C1}\label{eq:constraint_C1} \\
&\quad y_{j,k}(t) \in \{0,1\}, \quad \forall j \in \mathcal{J}, \forall k \in \mathcal{K} \tag{C2}\label{eq:constraint_C2} \\
&\quad z_{i,k}(t) \in \{0,1,\ldots,L_i\}, \quad \forall i \in \mathcal{I}, \forall k \in \mathcal{K} \tag{C3}\label{eq:constraint_C3} \\
&\quad \sum_{j=1}^{J} y_{j,k}(t) = 1, \quad \forall k \in \mathcal{K} \tag{C4}\label{eq:constraint_C4} \\
&\quad \sum_{k=1}^{K} f_{j,k}(t) \leq f_j, \quad \forall j \in \mathcal{J} \tag{C5}\label{eq:constraint_C5} \\
&\quad \sum_{k=1}^{K} B_{j,k}(t) \leq B_j, \quad \forall j \in \mathcal{J} \tag{C6}\label{eq:constraint_C6} \\
&\quad \sum_{i=1}^{I} x_{i,j}(t) \cdot D_i \leq S_j, \quad \forall j \in \mathcal{J} \tag{C7}\label{eq:constraint_C7} \\
&\quad \lim_{T \rightarrow \infty} \frac{1}{T} \sum_{t=0}^{T-1} \mathbb{E}[\bar{\tau}(t)] \leq \bar{\tau} \tag{C8}\label{eq:constraint_C8}
\end{align}
Here, constraint \eqref{eq:constraint_C4} ensures that each user is associated with exactly one server. Constraints \eqref{eq:constraint_C5} and \eqref{eq:constraint_C6} enforce the computation and bandwidth capacity limits at each server, and \eqref{eq:constraint_C7} guarantees that the total storage occupied by the deployed models does not exceed the server's storage capacity. Finally, \eqref{eq:constraint_C8} imposes the long-term average constraint on the system-wide per-user delay.

\section{Algorithm Design}\label{sec:algorithm}

Problem $\mathcal{P}$ in \eqref{eq:optimization_problem} constitutes an NP-hard, non-convex mixed-integer program with decisions coupled across $J$ servers and $K$ users. Its dynamic nature renders traditional optimization methods unsuitable. Although DRL is well-suited for such high-dimensional problems, standard DRL algorithms lack explicit mechanisms to enforce long-term average delay constraints as specified in \eqref{eq:constraint_C8}, often resulting in policies that violate delay thresholds and fail to provide reliable QoS guarantees.

To address these challenges, we propose HC-MAPPO-L, a hierarchical MARL algorithm that decomposes Problem~$\mathcal{P}$ into three decision layers aligned with their intrinsic timescales: (i) a deployment layer for strategic model caching at slow timescales, (ii) an association--partitioning layer for per-request user--server association and DNN partitioning, and (iii) an allocation layer for real-time resource management. To rigorously enforce the delay constraint, HC-MAPPO-L integrates Lagrangian relaxation into the MAPPO algorithm, a robust on-policy MARL algorithm recognized for its training stability and strong cooperative performance. The Lagrangian mechanism dynamically penalizes constraint violations, while the CTDE paradigm ensures effective coordination among agents. The following subsections elaborate on the framework's design, beginning with the hierarchical agent architecture and then detailing the key algorithmic components\footnote{For notational clarity: parenthetical subscripts denote environment time (e.g., $a_k(t)$), numeric subscripts index trajectory samples (e.g., $a_{k,t}$), and superscripts denote training iterations (e.g., $\theta^t$).}.

\subsection{Hierarchical Agent Architecture Design}

Based on the three-layer structure, we formulate the decision-making process using heterogeneous agent policies. The deployment and allocation layers operate as standard Decentralized Partially Observable Markov Decision Processes (Dec-POMDPs). In contrast, the association–partitioning layer is formulated as a Decentralized Partially Observable Constrained Markov Decision Process (Dec-POCMDP), wherein each user agent must satisfy the long-term delay constraint. To ensure tractability, we enforce a stricter per-agent delay constraint, which serves as a sufficient condition for satisfying the global objective. This distributed constraint handling via MAPPO-Lagrangian constitutes the core of our method.

\textbf{Observation Space:} Under partial observability, each agent's observation space is tailored to its specific decision-making role. Agent policies (i.e., actors) utilize local observations for decentralized execution, while their value functions (i.e., critics) are trained with access to augmented global information, thereby enabling enhanced coordination.

For the deployment agent on server $j$, the local observation supports its auto-regressive policy by tracking service popularity:
\begin{equation}
\label{eq:deploy_obs_local}
\mathbf{o}_j^{\mathrm{deploc}}(t) = \bigl[\, \mathbf{r}(t-{\scriptstyle\Delta}T, t),\; \mathbf{h}_j(t-{\scriptstyle\Delta}T, t),\; \mathbf{x}_j^{st} \bigr] \in \mathbb{R}^{3I}
\end{equation}
where $\mathbf{r}(\cdot)$ represents system-wide request counts (i.e., global popularity), $\mathbf{h}_j(\cdot)$ denotes requests to server $j$ (local demand), and $\mathbf{x}_j^{st}$ captures the server's current deployment state during auto-regressive steps (algorithm details are in Section~\ref{subsec:algorithm}).

During centralized training, the critic for each deployment agent $j$ uses an augmented observation $\mathbf{o}_j^{\mathrm{depglob}}(t)$. This global view enables coordinated learning without full joint-state dimensionality. The augmented observation is:
\begin{equation}
\label{eq:deploy_obs_global}
\mathbf{o}_j^{\mathrm{depglob}}(t) = \bigl[\, \mathbf{o}_j^{\mathrm{deploc}}(t),\; \mathrm{vec}(\mathbf{X}(t-{\scriptstyle\Delta}T)) \bigr] \in \mathbb{R}^{3I+IJ}
\end{equation}
where $\mathrm{vec}(\mathbf{X}(t-{\scriptstyle\Delta}T))$ is the vectorized system-wide deployment matrix from the previous deployment step. This global view allows the critic to account for inter-server dependencies, guiding actors away from suboptimal homogeneous strategies (e.g., all servers caching only the most popular models).

For user agent $k$, its local observation incorporates individual request details and the system-wide deployment status:
\begin{equation}
\mathbf{o}_k^{\mathrm{usrloc}}(t) = \bigl[ \mathbf{e}_k^{\mathrm{mod}}, \mathbf{e}_k^{\mathrm{inp}}, \mathrm{vec}(\mathbf{X}(t)) \bigr] \in \mathbb{R}^{d_m+d_s+IJ}
\end{equation}
where $\mathbf{e}_k^{\mathrm{mod}}$ and $\mathbf{e}_k^{\mathrm{inp}}$ are embedding vectors for the requested model and input data size of user $k$, with dimensions $d_m$ and $d_s$ respectively.

The global observation for the centralized critic aggregates request information from all users to learn a value function capable of resolving resource conflicts:
\begin{equation}
\mathbf{o}^{\mathrm{usrglob}}(t) = \bigl[\, \mathbf{n}^{\mathrm{mod}}(t),\; \mathbf{n}^{\mathrm{inp}}(t),\; \mathrm{vec}(\mathbf{X}(t)) \bigr] \in \mathbb{R}^{2K+IJ}
\end{equation}
Here, $\mathbf{n}^{\mathrm{mod}}(t)$ and $\mathbf{n}^{\mathrm{inp}}(t)$ denote the model IDs and input sizes of all $K$ users.

Finally, the allocation agent on each server $j$ operates autonomously, managing resources solely for its associated users. Its observation supports an attention-based policy with local request information:
\begin{equation}
\mathbf{o}_j^{\mathrm{alloc}}(t) = \bigl[\, \mathbf{n}_j^{\mathrm{mod}}(t),\; \mathbf{n}_j^{\mathrm{inp}}(t),\; \mathbf{n}_j^{\mathrm{spl}}(t) \bigr] \in \mathbb{R}^{3K}
\end{equation}
where the vectors encapsulate the service IDs, input sizes, and partitioning points for all users, masked such that only entries for users associated with server $j$ are non-zero. This structure enables the attention mechanism to dynamically focus on active users while preserving a fixed input dimension.

\textbf{Action Space:} The action space of each agent is customized according to its designated role.

The deployment agent on server $j$ determines its action every ${\scriptstyle\Delta}T$ time slots. Its action $\mathcal{M}_j^{\mathrm{dep}}(t)$ is a variable-length macro-action representing the deployment decision:
\begin{equation}
\mathcal{M}_j^{\mathrm{dep}}(t) = (a_1, a_2, \ldots, a_L)
\end{equation}
where each $a_n$ ($n=1,\ldots,L$) is a selected model index from $\mathcal{I}$, and $L \leq I$ is the number of models deployed (determined by storage capacity). The macro-action determines the binary deployment vector $\{x_{i,j}(t)\}_{i\in\mathcal{I}} \in \{0,1\}^{I}$.

The user agent $k$ makes decisions at every time slot. Its action $a_k^{\mathrm{usr}}(t)$ combines server association and model partitioning:
\begin{equation}
a_k^{\mathrm{usr}}(t) = (\sigma_k(t), z_{i,k}(t)) \in \{1,2,\ldots,J\} \times \{0,1,\ldots,L_i\}
\end{equation}
where $\sigma_k(t)$ is the selected server index, determining $\{y_{j,k}(t)\}_{j\in\mathcal{J}}$.

The allocation agent on server $j$ also operates per time slot. Its action $a_j^{\mathrm{alloc}}(t)$ specifies computational and bandwidth resources allocated to all users:
\begin{equation}
a_j^{\mathrm{alloc}}(t) = \bigl( \{f_{j,k}(t)\}_{k\in\mathcal{K}},\; \{B_{j,k}(t)\}_{k\in\mathcal{K}} \bigr)
\end{equation}
This high-dimensional continuous action is generated by an attention-based policy, which outputs normalized weights to allocate resource fractions across users, as further elaborated in Section~\ref{subsec:algorithm}.

\textbf{Reward Function:} To guide the learning process, we design distinct reward functions tailored to the role of each agent.
For deployment agents, a reward is computed every ${\scriptstyle\Delta}T$ time slots to balance service hit rate against model migration cost. The cumulative hit count during the deployment interval is defined as:
\begin{equation}\label{eq:hit_count}
H_j(t) = \sum_{t'=t-{\scriptstyle\Delta}T}^{t-1} \sum_{k \in \mathcal{U}_j(t')} \sum_{i \in \mathcal{I}} d_{k,i}(t') \cdot x_{i,j}(t')
\end{equation}
where $\mathcal{U}_j(t') = \{k \mid y_{j,k}(t') = 1\}$ is the set of users associated with server $j$, and $d_{k,i}(t')$ indicates if user $k$ requests model $i$.

The migration cost, $C_j^{\mathrm{mig}}(t)$, is the overhead of fetching new models:
\begin{equation}\label{eq:migration_cost}
C_j^{\mathrm{mig}}(t) = \sum_{i \in \mathcal{I}} x_{i,j}(t)\bigl(1 - x_{i,j}(t-{\scriptstyle\Delta}T)\bigr) \cdot \frac{D_i}{R^{C2E}}
\end{equation}
where the term $x_{i,j}(t)\bigl(1 - x_{i,j}(t-{\scriptstyle\Delta}T)\bigr)$ equals 1 if model $i$ is newly deployed.

The resulting reward for the deployment agent is:
\begin{equation}\label{eq:deploy_reward}
R_j^{\mathrm{dep}}(t) = \mu_{\mathrm{hit}} \, H_j(t) - \mu_{\mathrm{mig}} \, C_j^{\mathrm{mig}}(t)
\end{equation}
where $\mu_{\mathrm{hit}}, \mu_{\mathrm{mig}} > 0$ are weighting coefficients.

For user agents, the per-slot reward combines privacy cost, energy consumption, and delay violations:
\begin{equation}
R_k^{\mathrm{usr}}(t) =
\begin{cases}
-\big(\mu_1 C_k^{\mathrm{priv}}(t) + \mu_2 e_k(t) \\
\quad\;\; + \mu_3 [\tau_k(t)-\bar{\tau}]_+\big), & \text{if available,} \\
R^{\mathrm{fail}}, & \text{otherwise}
\end{cases}
\end{equation}
where $[x]_+ \triangleq \max(x,0)$ and $\mu_3>0$. The failure reward $R^{\mathrm{fail}}$ is applied for cache misses (i.e., $x_{i,j}(t) \cdot y_{j,k}(t) \neq 1$) to discourage infeasible associations. This immediate penalty complements the Lagrangian dual variable, which uses the raw delay $\tau_k(t)$ (from Eq.~\eqref{eq:total_delay}) to enforce the long-term constraint, while the positive-part term provides dense step-wise guidance.

For allocation agents, the reward is the negative average delay of its associated users, which its actions most directly influence:
\begin{equation}
R_j^{\mathrm{alloc}}(t) = -\frac{1}{|\mathcal{U}_j(t)|}\sum_{k \in \mathcal{U}_j(t)} \tau_k(t)
\end{equation}

\textbf{Constraint Cost Function:}
To handle the delay constraint, we define an instantaneous cost for each user agent equal to its end-to-end delay. This cost, $c_k(s_k(t),a_k(t))=\tau_k(t)$, drives the Lagrangian update. While the global objective is~\eqref{eq:constraint_C8}, our decentralized algorithm ensures each user agent $k$ satisfies a local, stricter condition:
\begin{equation}\label{eq:per_agent_constraint}
\lim_{T\rightarrow\infty}\frac{1}{T}\sum_{t=0}^{T-1}\mathbb{E}\!\left[c_k(s_k(t),a_k(t))\right]\le\bar{\tau}
\end{equation}
Satisfying this for all users is sufficient to guarantee the system-wide constraint.

\subsection{HC-MAPPO-L Algorithm Design}\label{subsec:algorithm}

\begin{figure*}[t]
\centering
\includegraphics[width=6.0in]{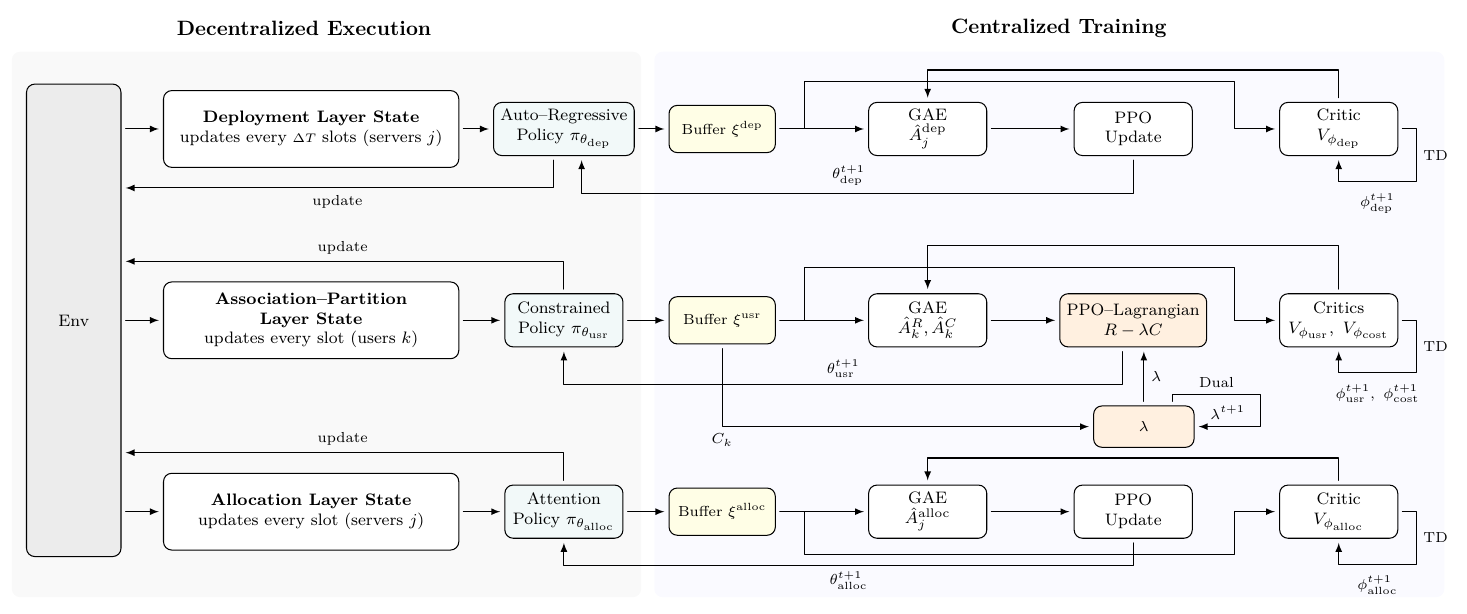}
\caption{The architecture of HC--MAPPO--L algorithm.}
\label{fig:hc_mappo_l_arch}
\end{figure*}

Addressing the hierarchical problem, we develop layer-specific policies for three core challenges: the combinatorial deployment action space, long-term delay constraint satisfaction, and resource allocation under dynamic user sets. HC-MAPPO-L integrates an auto-regressive deployment policy, Lagrangian constraint handling, an attention-based allocation policy, and a hierarchical CTDE training framework. The overall architecture is depicted in Fig.~\ref{fig:hc_mappo_l_arch}.

\subsubsection{Auto--Regressive--Based Model Deployment Policy}
\label{sec:ar_deploy_policy}

The deployment decision space for each server is $2^{|\mathcal{I}|}$, which is intractable for large model libraries. To manage this combinatorial complexity, we design an auto-regressive policy that decomposes the joint decision into a sequence of individual model selections. The deployment agent initializes an indicator vector $\mathbf{x}_j^{st}=\mathbf{0}$ and sequentially sets its entries. At each step $n$, the policy conditions on the observation $\mathbf{o}_j^{\mathrm{deploc}}(t)$ and outputs a probability distribution over $\mathcal{I}$, masked to enforce two constraints: (i) previously selected models are excluded, and (ii) models exceeding remaining storage are excluded, guaranteeing~\eqref{eq:constraint_C7}. The process terminates when no feasible models remain, reducing the exponential search to a polynomial-time procedure.

This deployment decision is a variable-length \emph{macro-action} $\mathcal{M}_j^{\mathrm{dep}}=(a_1,a_2,\ldots,a_L)$, and the joint probability factorizes as:
\begin{equation}
\pi_{\theta_{\mathrm{dep}}}\!\left(\mathcal{M}_j^{\mathrm{dep}}\mid s_j\right)=\prod_{n=1}^{L}\pi_{\theta_{\mathrm{dep}}}\!\bigl(a_n\mid h_n\bigr),
\end{equation}

\noindent where $h_n$ is the hidden state of a GRU that encodes the initial observation and previously selected models.

We incorporate two architectural enhancements. First, \emph{state augmentation} via concatenating the current deployment indicator $\mathbf{x}_j^{st}$ provides the policy a fine-grained view of remaining capacity. Second, an \emph{informed baseline} is created by having the critic share the GRU encoder to produce step-wise value estimates $V(h_n;\phi_{\mathrm{dep}})$. The value of the macro-action is the mean of valid step values:
\begin{equation}
V(\mathcal{M}_j^{\mathrm{dep}})=\frac{1}{L}\sum_{n=1}^{L}V\bigl(h_n;\phi_{\mathrm{dep}}\bigr).
\end{equation}

\noindent This baseline more accurately predicts the final reward, reducing advantage estimate variance and improving sample efficiency.

During training, the sequence is treated as one action. The sequence-level log-probability and advantage are:
\begin{align}
\log\pi_{\theta_{\mathrm{dep}}}(\mathcal{M}_j^{\mathrm{dep}}) &= \sum_{n=1}^{L}\log\pi_{\theta_{\mathrm{dep}}}(a_n\mid h_n)\\
\hat{A}_j^{\mathrm{dep}} &= \operatorname{GAE}\bigl(R(\mathcal{M}_j^{\mathrm{dep}}), V(\mathcal{M}_j^{\mathrm{dep}})\bigr)
\end{align}
where $R(\mathcal{M}_j^{\mathrm{dep}})$ is the total reward $R_j^{\mathrm{dep}}(t)$ from~\eqref{eq:deploy_reward}. The standard PPO loss is then applied. An illustration of this procedure is in Appendix~A.

\subsubsection{Constrained Optimization via Lagrangian Method}

A primary challenge is ensuring policies satisfy the delay constraint (\ref{eq:constraint_C8}). Standard MARL algorithms are ill-suited for such constraints. To address this, we integrate Lagrangian relaxation into the MAPPO framework. This method transforms the constrained problem into a saddle-point problem solved via alternating optimization. For the delay constraint, we introduce a shared non-negative Lagrange multiplier $\lambda \ge 0$ that dynamically adjusts based on constraint violations, acting as an adaptive penalty.

The Lagrangian for the user agents' problem is:
\begin{equation}\label{eq:lagrangian_objective}
\mathcal{L}_{\mathrm{usr}}(\pi_{\theta_{\mathrm{usr}}}, \lambda) = J_{\mathrm{usr}}^R(\pi_{\theta_{\mathrm{usr}}}) - \lambda \left(J_{\mathrm{usr}}^C(\pi_{\theta_{\mathrm{usr}}}) - \bar{J}^C\right)
\end{equation}
where $J_{\mathrm{usr}}^R$ and $J_{\mathrm{usr}}^C$ are the infinite-horizon discounted cumulative reward and delay cost, and $\bar{J}^C$ is the maximum tolerable cumulative delay. User agent policies are trained to maximize this objective. Concurrently, $\lambda$ is updated via projected gradient ascent:
\begin{equation}\label{eq:dual_update}
\lambda^{t+1} = \max\left(0, \lambda^t + \alpha_{\lambda} \cdot \left(\hat{J}_{\mathrm{usr}}^C - \bar{J}^C\right)\right)
\end{equation}
where $\alpha_{\lambda} > 0$ is the dual learning rate. This update increases $\lambda$ when the delay cost exceeds the threshold, steering the policy toward lower-delay actions, and decreases it otherwise, allowing the policy to prioritize rewards.

\subsubsection{Attention--Based Resource Allocation Policy}
\label{sec:attention_policy}

The dynamic number and characteristics of associated users challenge server-side resource allocation. We design an attention-based policy that adapts resource distribution to user requirements. The policy first encodes the server's observation vector $\mathbf{o}_j^{\mathrm{alloc}}(t)$ into a unified context vector, $\mathbf{g}_j(t)$, which serves as the attention query.

The mechanism uses a query–key scheme. For each associated user $k \in \mathcal{U}_j(t)$, a key vector $\boldsymbol{\kappa}_k = \mathrm{MLP}_{\mathrm{key}}([\mathbf{e}_k^{\mathrm{mod}}; \mathbf{e}_k^{\mathrm{inp}}; \mathbf{e}_k^{\mathrm{spl}}])$ is constructed from model, input size, and partitioning point embeddings. The raw attention score is computed as $\mathbf{g}_j(t)^T \boldsymbol{\kappa}_k / \sqrt{d_h}$, where $d_h$ denotes the key dimension, and normalized via softmax to produce allocation weights.

To model distinct allocation priorities, the policy employs two parallel attention branches for computation and bandwidth resources. Each branch generates a specialized query ($\mathbf{q}_j^{\mathrm{comp}}(t)$, $\mathbf{q}_j^{\mathrm{band}}(t)$) and key set ($\{\boldsymbol{\kappa}_k^{\mathrm{comp}}\}$, $\{\boldsymbol{\kappa}_k^{\mathrm{band}}\}$) from the shared context and base user features, respectively. The resulting weight distributions are concatenated to form the action: $a_j^{\mathrm{alloc}}(t) = [\mathbf{w}_j^{\mathrm{comp}}(t); \mathbf{w}_j^{\mathrm{band}}(t)]$. This process is detailed with a diagram in Appendix~A.

\subsubsection{Hierarchical CTDE Training Framework}
The HC-MAPPO-L algorithm is trained using a CTDE paradigm tailored to the three-layer architecture. This accommodates heterogeneous agent types and update intervals, leveraging centralized training for coordination while maintaining decentralized execution.

The implementation uses layer-specific network architectures with parameter sharing within each layer to balance efficiency and performance. Deployment agents use auto-regressive policies ($\pi_{\theta_{\mathrm{dep}}}$) and critics ($V_{\phi_{\mathrm{dep}}}$); user agents use policies ($\pi_{\theta_{\mathrm{usr}}}$) with local observations, paired with centralized value ($V_{\phi_{\mathrm{usr}}}$) and cost ($V_{\phi_{\mathrm{cost}}}$) critics that use global observations; allocation agents use attention-based policies ($\pi_{\theta_{\mathrm{alloc}}}$) and critics ($V_{\phi_{\mathrm{alloc}}}$). Despite parameter sharing, the framework generates distinct trajectories for each agent based on its unique state-action history.

For deployment agents (updating every ${\scriptstyle\Delta}T$ slots), trajectories are:
\begin{equation}
\xi_j^{\mathrm{dep}} = \{(\mathbf{o}_{j,t}^{\mathrm{dep}}, \mathcal{M}_{j,t}^{\mathrm{dep}}, r_{j,t}^{\mathrm{dep}}, \mathbf{o}_{j,t+{\scriptstyle\Delta}T}^{\mathrm{dep}})\}_{t \in \{n {\scriptstyle\Delta}T \mid n \in \mathbb{N}\}}
\end{equation}
where $\mathcal{M}_{j,t}^{\mathrm{dep}}$ is the macro-action from Section~\ref{subsec:algorithm}.

For user agents (updating each slot), trajectories include the constraint cost:
\begin{equation}
\xi_k^{\mathrm{usr}} = \{(\mathbf{o}_{k,t}^{\mathrm{usr}}, a_{k,t}^{\mathrm{usr}}, r_{k,t}^{\mathrm{usr}}, c_{k,t}, \mathbf{o}_{k,t+1}^{\mathrm{usr}})\}_{t=0}^{T-1}
\end{equation}

For allocation agents (updating each slot), trajectories are:
\begin{equation}
\xi_j^{\mathrm{alloc}} = \{(\mathbf{o}_{j,t}^{\mathrm{alloc}}, a_{j,t}^{\mathrm{alloc}}, r_{j,t}^{\mathrm{alloc}}, \mathbf{o}_{j,t+1}^{\mathrm{alloc}})\}_{t=0}^{T-1}
\end{equation}

Based on these trajectories, the hierarchical training process uses synchronized data collection and layer-specific updates. The centralized training uses these individual trajectories to update shared parameters. For deployment agents, standard PPO updates are performed:
\begin{equation}
\theta_{\mathrm{dep}} \leftarrow \theta_{\mathrm{dep}} + \alpha_{\mathrm{dep}} \nabla_{\theta_{\mathrm{dep}}} \hat{J}_{\mathrm{dep}}^R(\pi_{\theta_{\mathrm{dep}}})
\end{equation}
where $\hat{J}_{\mathrm{dep}}^R$ is the PPO clipped surrogate objective~\cite{Schulman2017PPO}, estimated from deployment trajectories.

For user agents, the PPO-Lagrangian method optimizes the Lagrangian objective~\eqref{eq:lagrangian_objective}. The policy is updated via:
\begin{equation}
\theta_{\mathrm{usr}} \leftarrow \theta_{\mathrm{usr}} + \alpha_{\mathrm{usr}} \left[\nabla_{\theta_{\mathrm{usr}}} \hat{J}_{\mathrm{usr}}^R(\pi_{\theta_{\mathrm{usr}}}) - \lambda \nabla_{\theta_{\mathrm{usr}}} \hat{J}_{\mathrm{usr}}^C(\pi_{\theta_{\mathrm{usr}}})\right]
\end{equation}
where $\hat{J}_{\mathrm{usr}}^R$ and $\hat{J}_{\mathrm{usr}}^C$ are PPO objectives estimated from reward and cost advantage estimates $\hat{A}_k^R$ and $\hat{A}_k^C$, respectively. Concurrently, the Lagrange multiplier is updated via the dual ascent rule in Eq.~\eqref{eq:dual_update}. Allocation agents use the same PPO update mechanism as deployment agents.

All centralized critics are updated by minimizing the mean squared error loss over their respective trajectory datasets. The complete HC-MAPPO-L algorithm is detailed in Algorithm~\ref{alg:hc_mappo_lagrangian}.

\begin{algorithm}[t]
\caption{Hierarchical Constrained MAPPO-Lagrangian (HC-MAPPO-L) Algorithm}
\label{alg:hc_mappo_lagrangian}
\begin{algorithmic}[1]
\Require Initial parameters $\theta_{\mathrm{dep}}^0, \theta_{\mathrm{usr}}^0, \theta_{\mathrm{alloc}}^0$, critics $\phi_{\mathrm{dep}}^0, \phi_{\mathrm{usr}}^0, \phi_{\mathrm{alloc}}^0, \phi_{\mathrm{cost}}^0$, $\lambda^0 = 0$
\Ensure Converged policies $\pi_{\theta_{\mathrm{dep}}}^{*}, \pi_{\theta_{\mathrm{usr}}}^{*}, \pi_{\theta_{\mathrm{alloc}}}^{*}$
\State Initialize all parameters and replay buffers
\While{not converged}
    \State \textit{// Data Collection}
    \For{$N$ episodes}
        \If{$t \bmod \Delta T = 0$}
            \State Execute auto-regressive deployment: $\mathcal{M}_j^{\mathrm{dep}}(t) \sim \pi_{\theta_{\mathrm{dep}}}$ for all servers $j$
        \EndIf
        \State Execute user association and model partitioning: $a_k^{\mathrm{usr}}(t) \sim \pi_{\theta_{\mathrm{usr}}}$ for all users $k$
        \State Execute resource allocation: $a_j^{\mathrm{alloc}}(t) \sim \pi_{\theta_{\mathrm{alloc}}}$ for all servers $j$
        \State Store trajectories $\xi_j^{\mathrm{dep}}$, $\xi_k^{\mathrm{usr}}$, $\xi_j^{\mathrm{alloc}}$
    \EndFor
    \State \textit{// Advantage Computation}
    \State Compute GAE advantages: $\hat{A}_j^{\mathrm{dep}}$, $\hat{A}_k^R$, $\hat{A}_k^C$, $\hat{A}_j^{\mathrm{alloc}}$
    \State \textit{// Lagrangian Update}
    \State $\hat{J}_{\mathrm{usr}}^C \leftarrow \frac{1}{NK}\sum_{\text{episodes}}\sum_k c_k$
    \State $\lambda \leftarrow \max(0, \lambda + \alpha_{\lambda} \cdot (\hat{J}_{\mathrm{usr}}^C - \bar{J}^C))$
    \State \textit{//Policy Updates}
    \State Update $\theta_{\mathrm{dep}}$ via PPO using $\hat{A}_j^{\mathrm{dep}}$
    \State Update $\theta_{\mathrm{usr}}$ via PPO-Lagrangian using $\hat{A}_k^R$, $\hat{A}_k^C$, $\lambda$
    \State Update $\theta_{\mathrm{alloc}}$ via PPO using $\hat{A}_j^{\mathrm{alloc}}$
    \State\textit{ // Critic Updates}
    \State Update critics $\phi_{\mathrm{dep}}, \phi_{\mathrm{usr}}, \phi_{\mathrm{alloc}}, \phi_{\mathrm{cost}}$ via MSE loss
\EndWhile
\end{algorithmic}
\end{algorithm}

\section{Experiments and Analysis}\label{sec:experiments}

\subsection{Experimental Design}
\label{sec:exp_setup}
\textbf{Experimental setup}: We evaluate HC-MAPPO-L through simulations with $J{=}10$ edge servers and $K{=}50$ users distributed over a 1000~m$\times$1000~m area, where servers are grid-placed and users randomly located~\cite{10609797}. The service pool includes nine base models (LeNet-7/9/12, ResNet-18/34/50, VGG-13/16/19)~\cite{11153065}, each instantiated as five services ($I{=}45$), with requests following a Zipf distribution ($s{=}0.8$)~\cite{Breslau1999Zipf}. Our evaluation examines three aspects: (i) convergence and cost-delay trade-offs under long-term delay constraints; (ii) scalability with number of users, edge servers, and service diversity; and (iii) effectiveness of Lagrangian constraint handling. Key parameters are provided in Table~\ref{tab:sim_params}. All base models were profiled to parameterize the DNN partitioning model, with detailed VGG-19 profiles available in the Appendix~B as an example.
\begin{table}[t]
\centering
\caption{Key Simulation Parameters}
\label{tab:sim_params}
\resizebox{0.98\columnwidth}{!}{%
\begin{tabular}{ll}
\toprule
\textbf{Parameter} & \textbf{Value} \\
\midrule
User compute capacity ($f_k$) & 10--100~GFLOPS (Uniform) \\
Edge compute capacity ($f_j$) & 500--2000~GFLOPS (Uniform) \\
User storage capacity & 1--2~GB (Uniform) \\
Edge storage capacity & 3--5~GB (Uniform) \\
Per-server bandwidth & 50--100~MHz (Uniform) \\
Server / User TX power & 30--43 / 20--30~dBm \\
Energy coefficient ($\epsilon_k$) & $10^{-11}$--$10^{-9}$~J/FLOP (Uniform)~\cite{Horowitz2014Energy} \\
Path-loss exponent / Shadowing & 3.5 / 8~dB \\
Reference path loss (1~m) & 30~dB \\
Carrier frequency & 3.5~GHz \\
Receiver noise figure & 6~dB \\
Cloud-to-edge download rate & 200--500~Mbps (Uniform) \\
Delay constraint ($\bar{\tau}$) & 3~seconds~\cite{10413648} \\
Deployment update interval ($\Delta T$) & 10~slots \\
User privacy preference ($Pre_k$) & 0.2--0.8 (Uniform) \\
Privacy coefficients & $\alpha_1=0.31, \alpha_2=1.88$ \\
Cost weights & $\mu_1=5,\ \mu_2=5$ \\
\bottomrule
\end{tabular}%
}
\end{table}
We implement all agent policies in PyTorch using orthogonal initialization and Tanh activations. The user policy employs a MAPPO architecture with decentralized actors and a centralized critic, both implemented as 256-dimensional MLPs. Deployment policies use GRU-based auto-regressive models with 256-dimensional hidden state, while allocation policies adopt attention-based networks (128-dim hidden size) with dual parallel heads. Training runs for 1000 iterations of 200-step data collection, with shared PPO parameters: learning rate $3\times10^{-4}$, $\gamma=0.99$, $\lambda_{\mathrm{GAE}}=0.95$, clipping coefficient 0.2, and entropy coefficients of 0.05 (user) and 0.25 (deployment) to encourage exploration. The Lagrange multiplier is initialized to $\lambda^0=0.01$ and updates with $\alpha_{\lambda}=0.01$ within $[0, 100]$\footnote{Code: \url{https://github.com/wanghong5233/HC-MAPPO-L}}.

\textbf{Baseline Selection}: For baseline algorithms, we consider two categories: heuristic algorithms commonly used in edge computing~\cite{Mach2017MECSurvey,Mao2017MECSurveyComm} and RL ablations. Heuristic methods include:
\begin{itemize}
\item \textbf{Local-Only}: All inference tasks are executed locally with channel-based association, equal-share allocation, and popularity-based deployment.
\item \textbf{Edge-Only}: All inference tasks are offloaded using identical association, allocation, and deployment strategies as Local-Only.
\item \textbf{Greedy}: Selects the strongest-channel server with requested service, chooses the deepest feasible partition point under equal-share delay estimation, and uses popularity-based deployment.
\end{itemize}

RL-based ablations systematically evaluate key components of our framework, including adaptations to alternative MARL paradigms like Independent PPO (IPPO)~\cite{NEURIPS2022_9c1535a0}:
\begin{itemize}
    \item \textbf{Heuristic-MAPPO-L}: Integrates our constrained user policy with heuristic resource management, e.g., equal-share resource allocation and Least Recently Used (LRU)-based model deployment to evaluate learned allocation and deployment.
    \item \textbf{H-IPPO}: Unconstrained independent PPO variant without centralized critic, optimizing for reward only.
    \item \textbf{HC-IPPO-L}: Constrained IPPO variant applying our Lagrangian mechanism to independent policies, assessing centralized training benefits.
    \item \textbf{H-MAPPO}: Unconstrained ablation removing the Lagrangian mechanism to quantify its contribution.
\end{itemize}

Overall, HC-MAPPO-L uniquely combines hierarchical MARL, Lagrangian-based delay constraint handling, and attention-based resource allocation, distinguishing it from the above baselines in terms of safety and efficiency.

\textbf{Evaluation Metrics}: We evaluate performance using five metrics: (i) average per-user energy consumption (J), (ii)~average privacy cost, and (iii)~average delay (s) for QoS assessment, (iv)~service success rate for reliability, and (v)~average user cost. All results are averaged over multiple random seeds.

\subsection{Results Analysis}

\begin{figure}[t]
    \centering
    \includegraphics[width=3.3in]{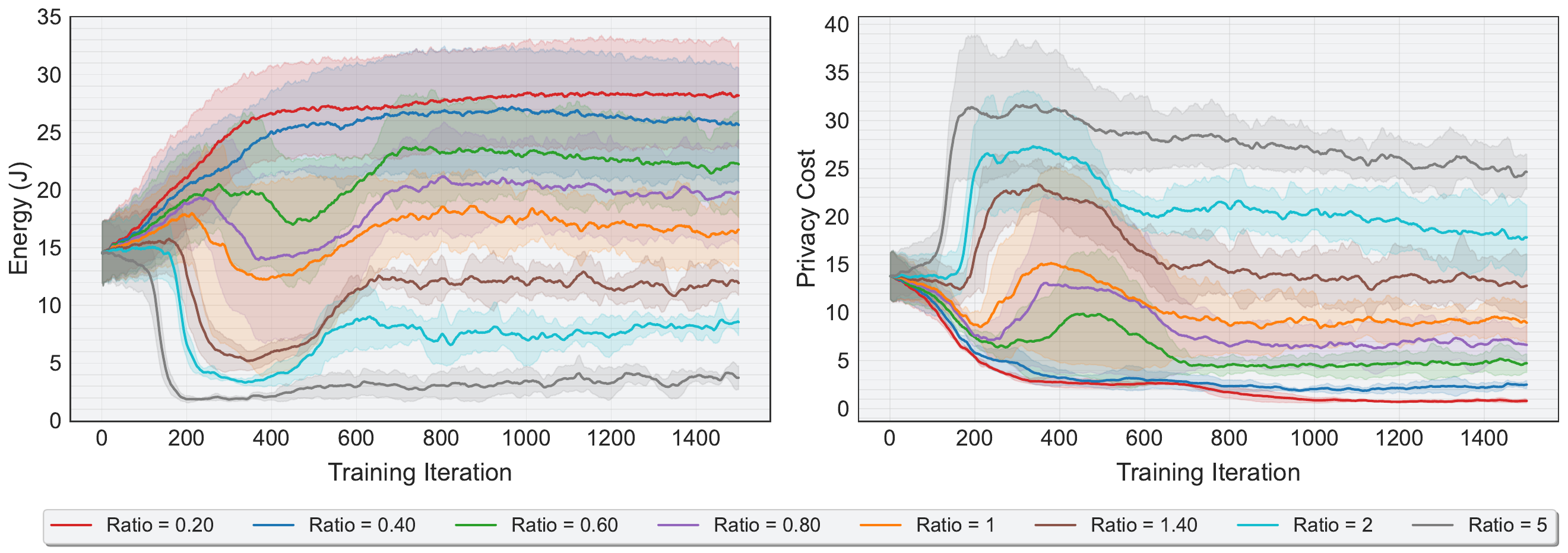}
    \caption{Evaluation of convergence performance versus energy--privacy weight ratios.}
    \label{fig:weight_convergence}
\end{figure}

\textbf{Energy--Privacy Trade-off Evaluation}: We first calibrate the weights to evaluate the trade-off for energy and privacy costs. Fixing the privacy weight $\mu_1=5$, we sweep the energy weight $\mu_2$ from 1 to 25, yielding an energy-to-privacy weight ratio $r\!=\!\mu_2/\mu_1\in[0.2,5]$. As shown in Fig.~\ref{fig:weight_convergence}, small ratios ($r<1$) prioritize privacy at the cost of higher energy, while large ratios ($r>1$) reduce energy at the expense of privacy. A ratio of $r=1$ ($\mu_1=5, \mu_2=5$) achieves a balanced trade-off, with stable per-user energy ($\approx$15--18~J) and privacy cost ($\approx$10--12) while satisfying the latency constraint. We thus adopt these weights for all subsequent experiments.

\begin{figure}[h]
    \centering
    \includegraphics[width=3.3in]{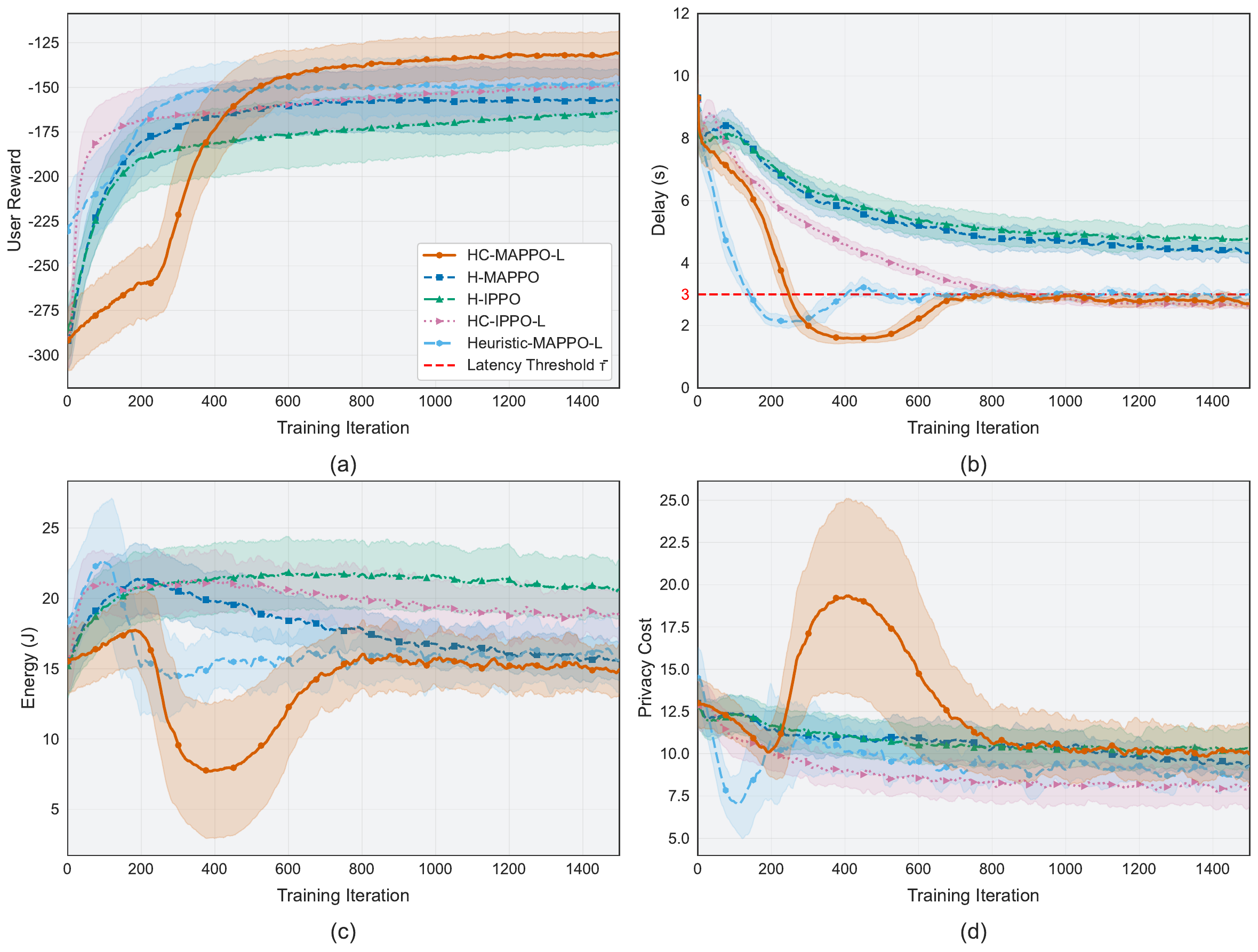}
    \caption{Evaluation of training convergence performance. All subfigures share the legend in (a), and the red dashed line in (b) indicates the latency constraint $\bar{\tau}=3$~s.}
    \label{fig:conv_cost_delay}
\end{figure}

\textbf{Convergence Performance Evaluation}: Figure~\ref{fig:conv_cost_delay} illustrates training convergence under the 3s delay constraint, revealing the critical importance of Lagrangian mechanisms for QoS assurance. All Lagrangian-based methods maintain delays below the required threshold: HC-MAPPO-L: 2.74 s, HC-IPPO-L: 2.65 s, Heuristic-MAPPO-L: 2.94 s, while unconstrained baselines exhibit significant violations: H-MAPPO: 4.38 s, H-IPPO: 4.76 s. This performance gap underscores the necessity of constrained optimization for reliable service delivery.

The Lagrangian mechanism enables this constraint satisfaction through dynamic policy adaptation. In HC-MAPPO-L, the multiplier guides learning through three phases: initial emphasis on privacy via deeper partitions, transitional shift toward latency reduction under penalty pressure, and final rebalancing once delays stabilize below the threshold. This responsive adaptation yields superior overall performance, with HC-MAPPO-L achieving the lowest user cost of 131.35 and energy consumption 14.87 J, while maintaining competitive privacy protection with 10.09. The coordinated hierarchical approach proves essential for resolving resource contention and simultaneously optimizing multiple objectives under strict QoS requirements.

\begin{figure*}[t]
    \centering
    \includegraphics[width=6.6in]{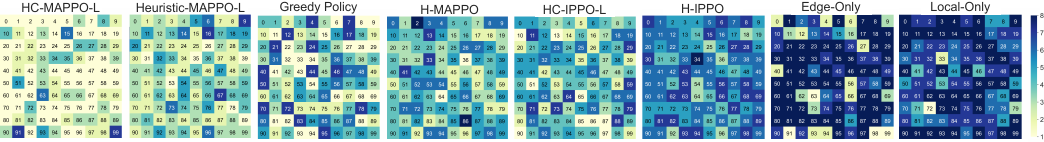}
    \caption{Evaluation of user cost distribution. The right-hand color bar range $[1, 8]$ represents the normalized cost rank, where 1 (8) corresponds to the lowest (highest) user cost.}
    \label{fig:heatmap_unified_cost_rank}
\end{figure*}

\textbf{User Cost Distribution Evaluation}: After verifying overall convergence, Fig.~\ref{fig:heatmap_unified_cost_rank} assesses system fairness via user-level cost visualization, where lighter colors denote lower costs. Specifically, HC-MAPPO-L achieves markedly superior balance, with about 54\% of users incurring the lowest costs and exhibiting a uniformly light distribution, indicating minimal deviation from the average cost. In contrast, heuristic methods show dark clusters representing disadvantaged users, while unconstrained RL methods produce dark bands due to reward-only optimization. Overall, these results demonstrate that HC-MAPPO-L effectively balances system efficiency and user fairness, by tolerating locally suboptimal outcomes for certain users, it preserves global fairness while maintaining a low average cost and preventing excessive burden on specific participants.

\begin{figure}[t]
    \centering
    \includegraphics[width=3.3in]{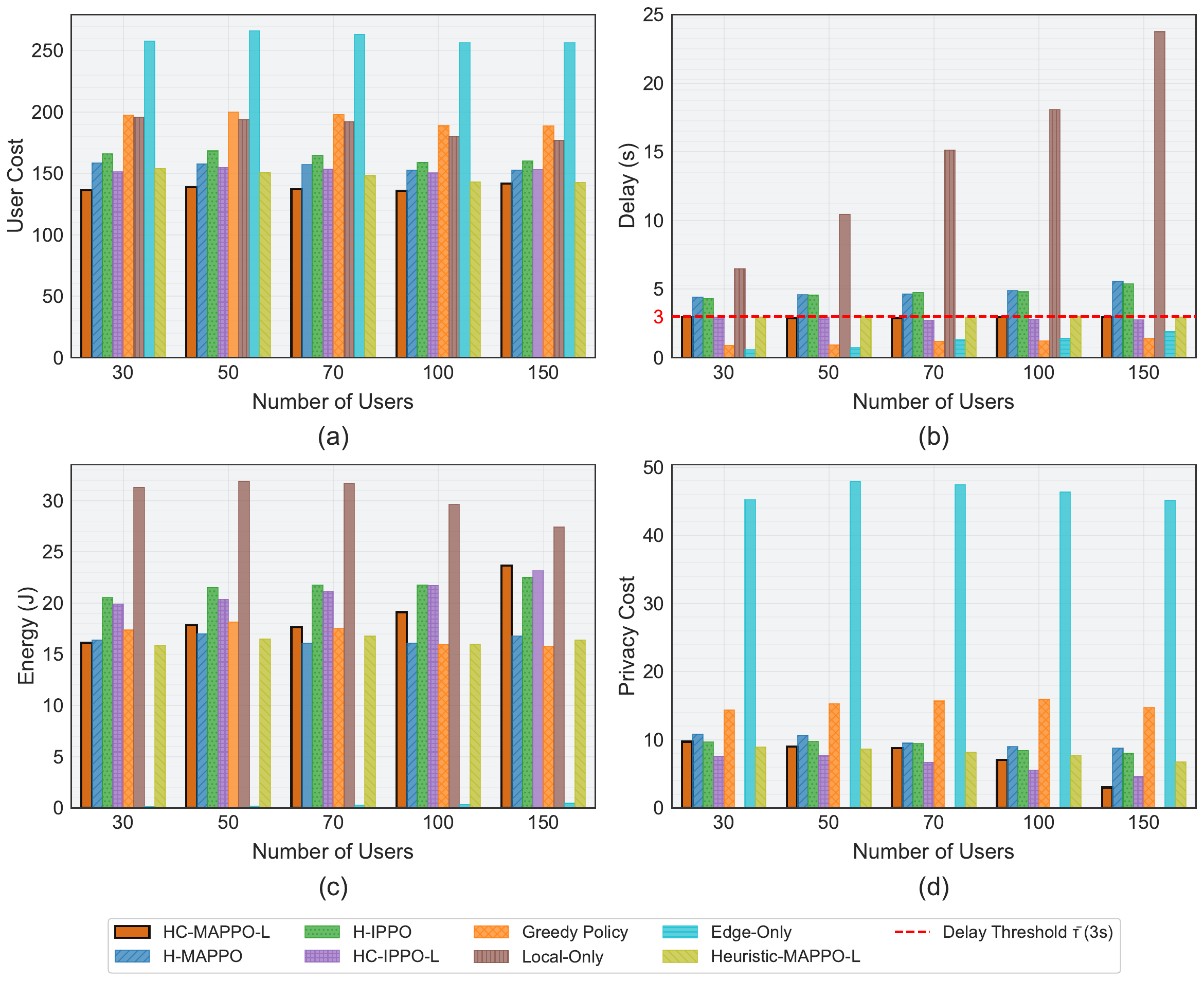}
    \caption{Evaluation of system performance under varying number of users.}
    \label{fig:sweep_users}
\end{figure}
\begin{figure}[t]
    \centering
    \includegraphics[width=3.3in]{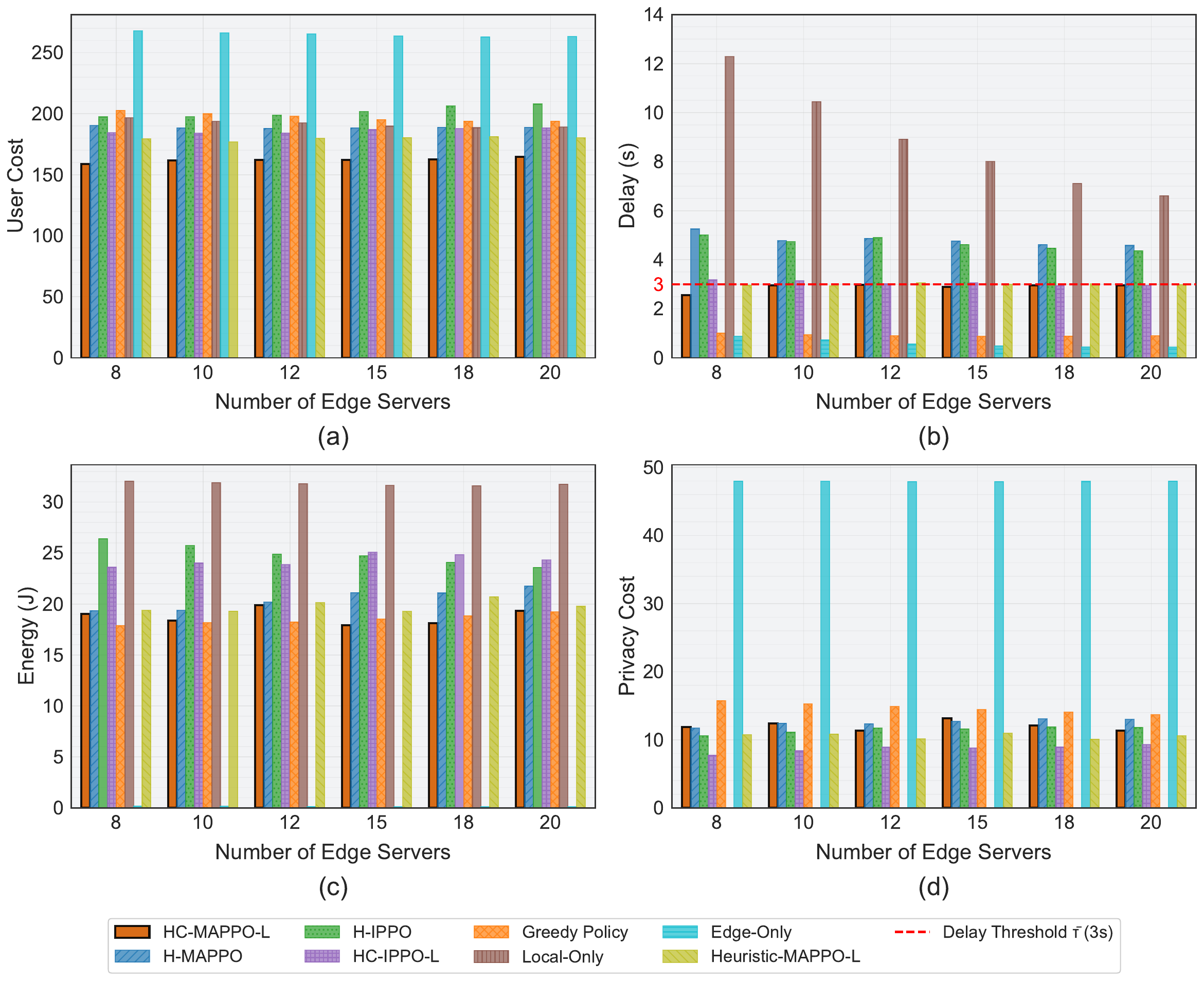}
    \caption{Evaluation of system performance under varying number of edge servers.}
    \label{fig:sweep_servers}
\end{figure}
\begin{figure}[t]
    \centering
    \includegraphics[width=3.3in]{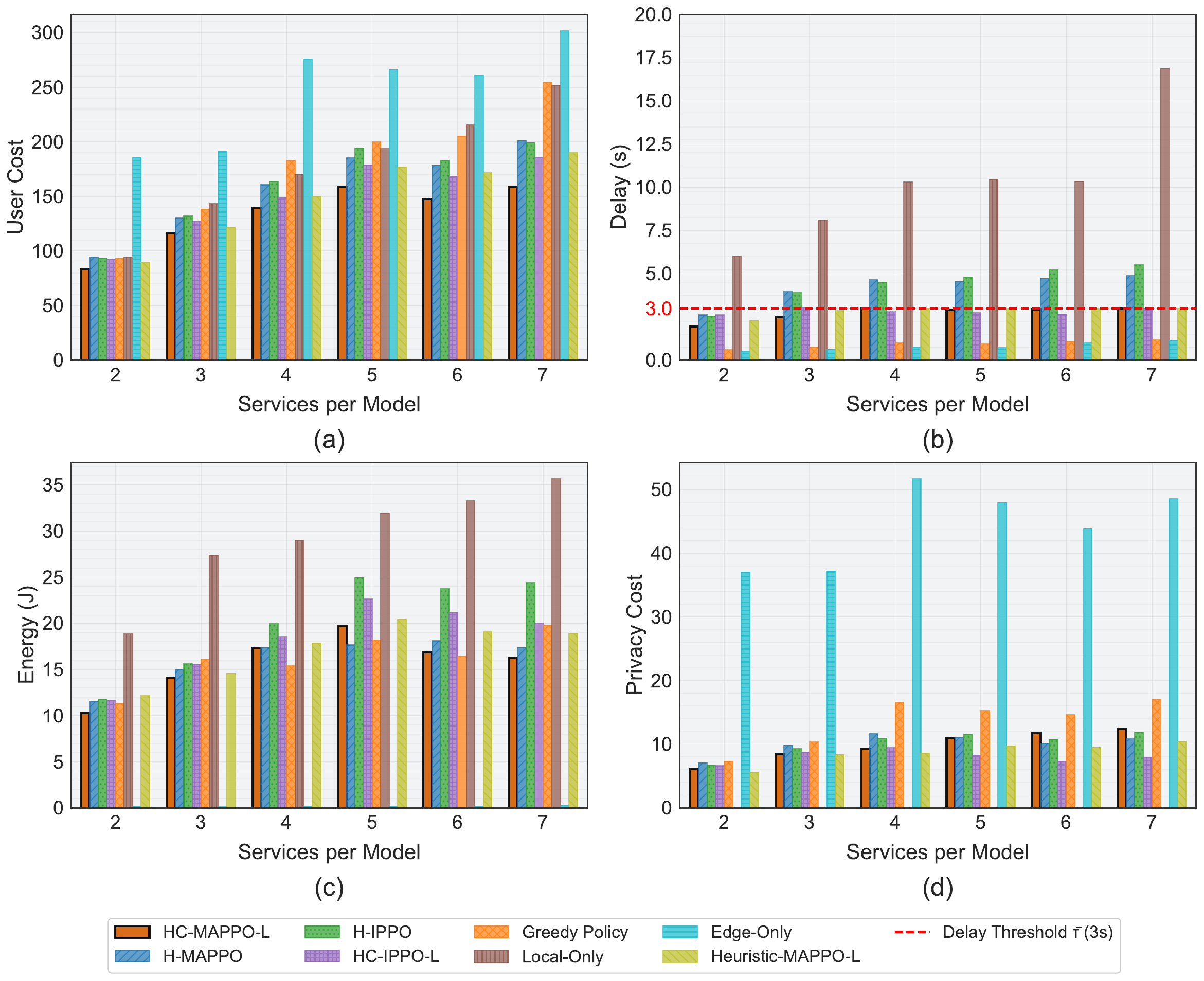}
    \caption{Evaluation of system performance under varying number of services.}
    \label{fig:scalability_service_per_model_bars}
\end{figure}
\textbf{Impact of Number of Edge Servers}: Fig.~\ref{fig:sweep_servers} evaluates scalability by varying the number of servers $J$ from 8 to 20. HC-MAPPO-L consistently yields the lowest user cost, outperforming Heuristic-MAPPO-L and unconstrained baselines by about 12\% and 21\%, respectively. Constrained methods satisfy the delay requirement ($<$3~s) under all settings, whereas unconstrained baselines incur persistent violations ($>$4.5~s) despite increased server availability, highlighting the importance of constraint-aware optimization. Moreover, HC-MAPPO-L maintains a stable energy--privacy trade-off (energy $\approx$19~J, privacy $\approx$11.5) across different scales, indicating that the hierarchical policy exploits additional edge resources without degrading QoS or privacy.

\textbf{Impact of Number of Edge Servers}: Fig.~\ref{fig:sweep_servers} evaluates scalability by varying server number $J$ from 8 to 20. HC-MAPPO-L consistently achieves the lowest user cost, outperforming Heuristic-MAPPO-L and unconstrained baselines by approximately 12\% and 21\%, respectively. While constrained methods maintain strict delay compliance ($<$3~s) across all settings, unconstrained baselines exhibit persistent violations ($>$4.5~s) even with increased server availability, underscoring the necessity of constraint-aware optimization. Furthermore, HC-MAPPO-L sustains a stable energy-privacy balance (energy $\approx$19~J, privacy $\approx$11.5) regardless of infrastructure scale, demonstrating that our hierarchical policy effectively leverages additional edge resources to minimize cost without compromising QoS or privacy.

\begin{figure}[t]
    \centering
    \includegraphics[width=3.3in]{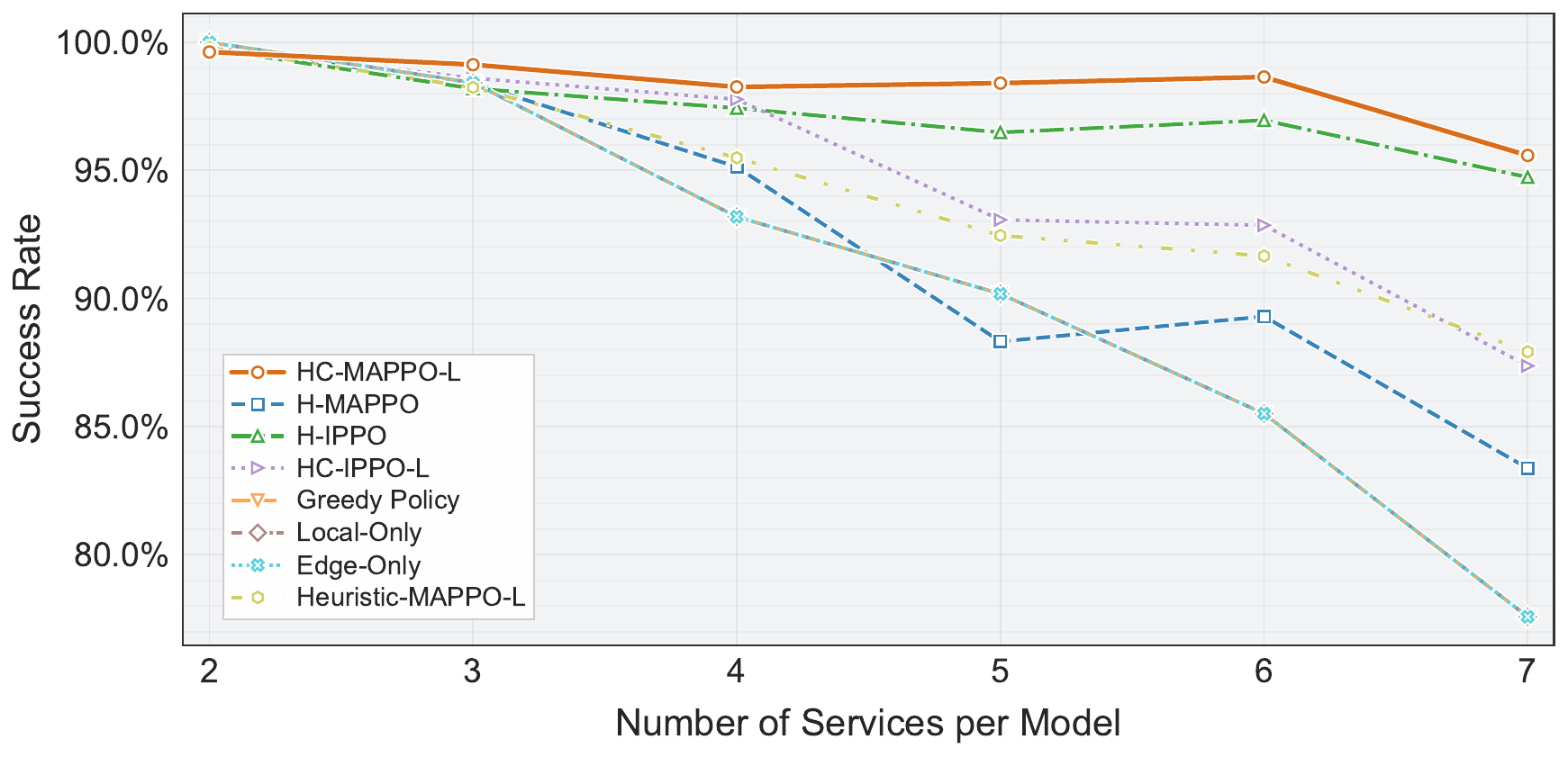}
    \caption{Service success rate versus services per model.}
    \label{fig:success_rate_services}
\end{figure}

\begin{figure}[t]
    \centering
    \includegraphics[width=3.3in]{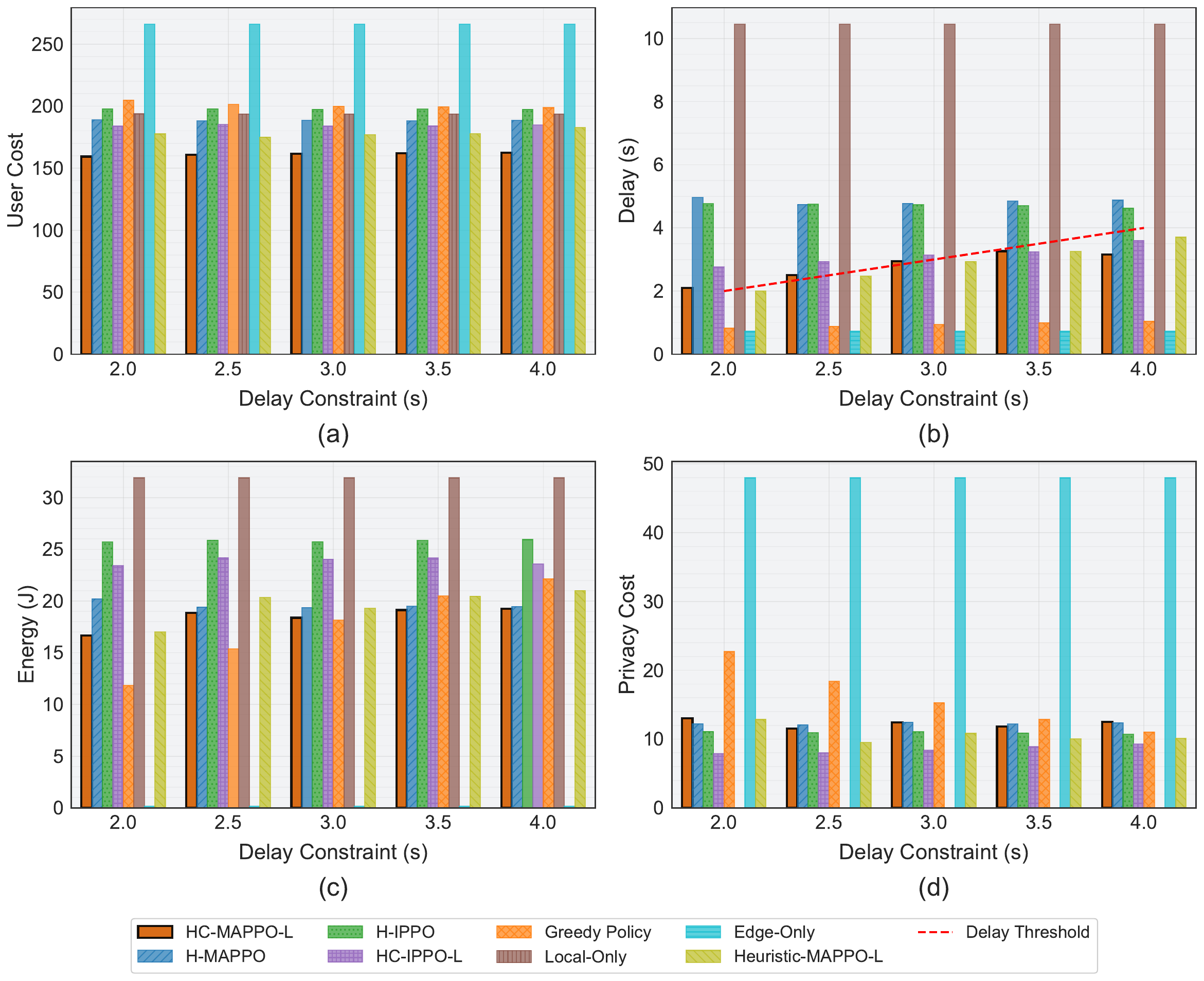}
    \caption{Evaluation of system performance under varying delay constraints.}
    \label{fig:sensitivity_delay_constraint_bars}
\end{figure}

\begin{figure}[t]
    \centering
    \includegraphics[width=3.3in]{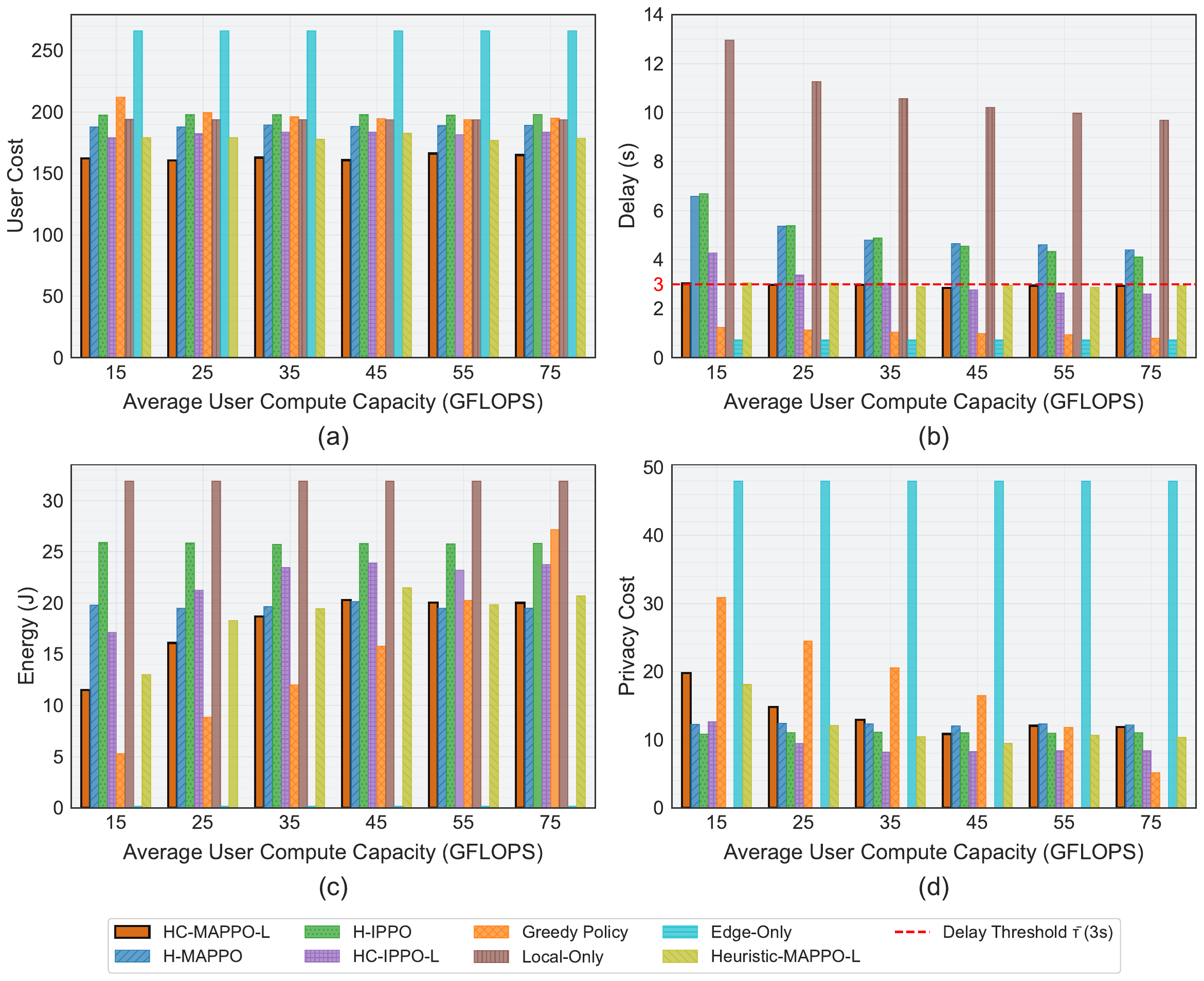}
    \caption{Evaluation of system performance under varying user computation capacities.}
    \label{fig:sensitivity_user_compute_bars}
\end{figure}
\begin{figure}[t]
    \centering
    \includegraphics[width=3.3in]{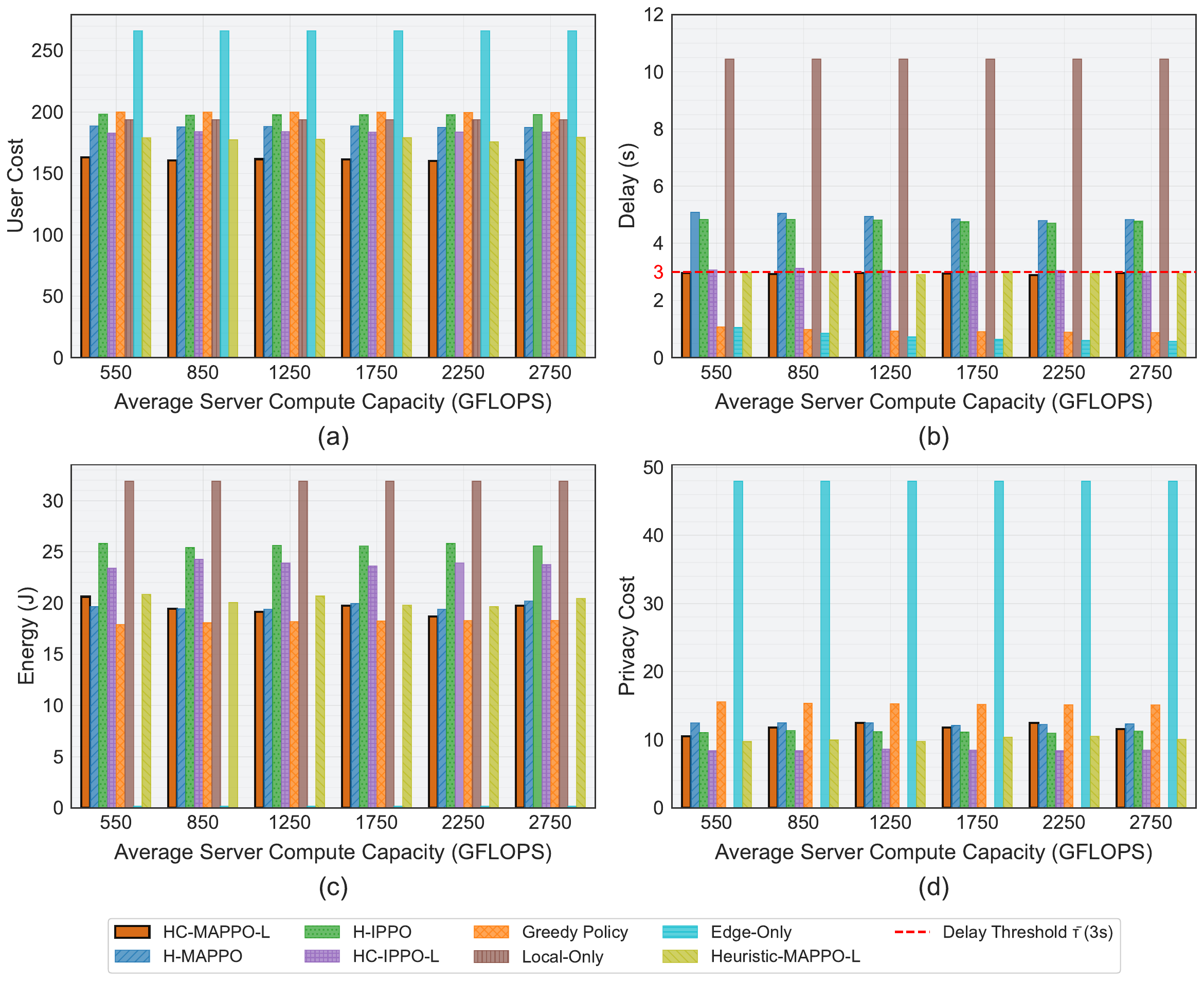}
    \caption{Evaluation of system performance under varying server computation capacities.}
    \label{fig:sensitivity_server_compute_bars}
\end{figure}

\begin{figure}[t]
    \centering
    \includegraphics[width=3.3in]{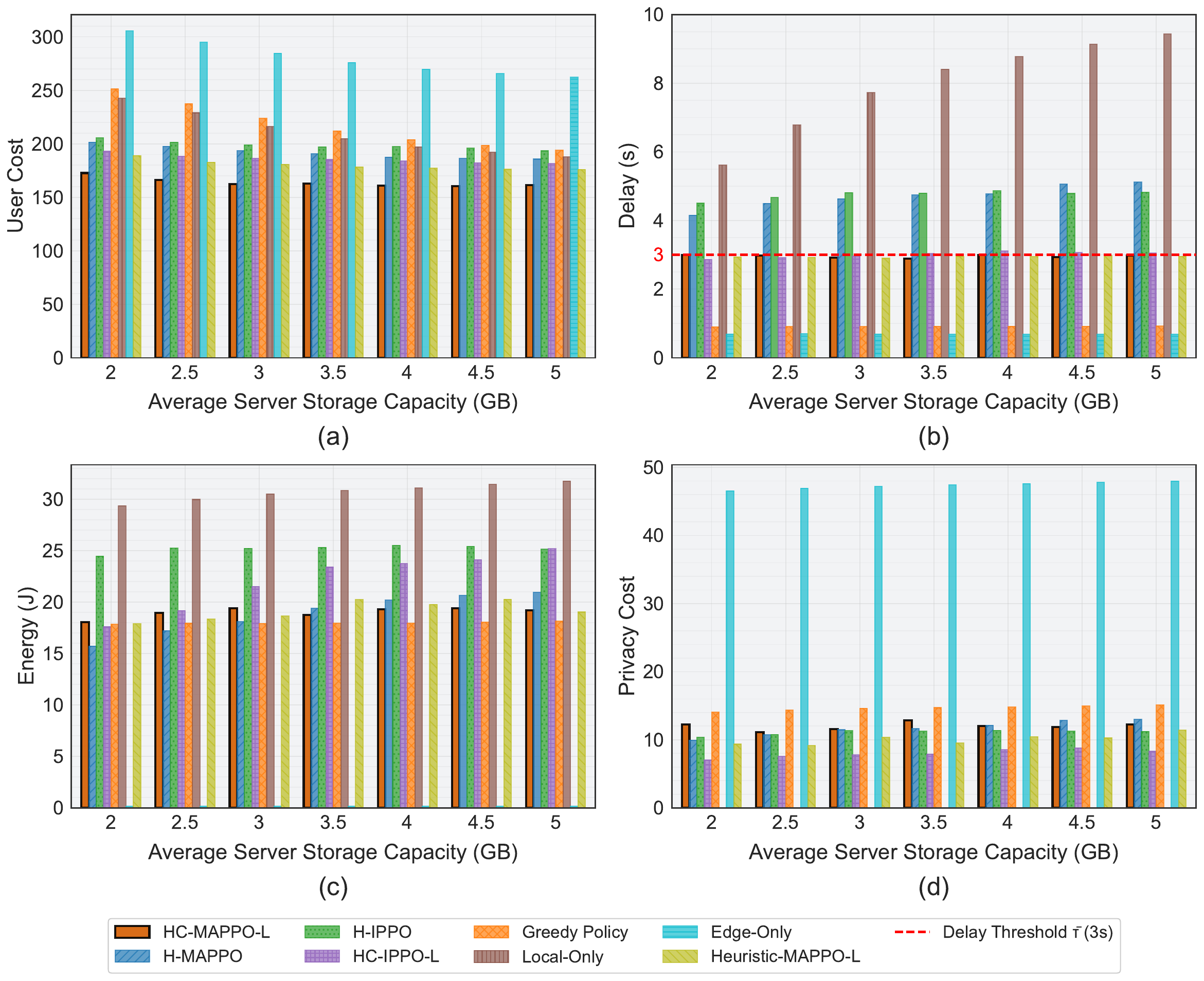}
    \caption{Evaluation of system performance under varying server storage capacity.}
    \label{fig:sensitivity_server_storage_bars}
\end{figure}

\textbf{Impact of Number of Services}: Fig.~\ref{fig:scalability_service_per_model_bars} evaluates system scalability under increasing service diversity. Fig.~\ref{fig:scalability_service_per_model_bars}(a) demonstrates HC-MAPPO-L's superior cost efficiency, achieving 158.38 at maximum diversity while outperforming alternatives by 14.4-37.8\%. Fig.~\ref{fig:scalability_service_per_model_bars}(b) confirms all constrained methods maintain delay compliance across diversity levels. Fig.~\ref{fig:scalability_service_per_model_bars}(c) and (d) reveal HC-MAPPO-L's balanced performance, maintaining energy consumption at 16-20 J and privacy costs at 6-12 through adaptive partitioning strategy.

Fig.~\ref{fig:success_rate_services} demonstrates deployment robustness under service diversity. HC-MAPPO-L achieves 96\% success rate at maximum diversity, exceeding alternatives by 8-18\%. This advantage stems from the auto-regressive policy's efficient cache utilization, ensuring reliable service availability despite increasing model variety and storage constraints.

\textbf{Impact of Delay Constraints}: Fig.~\ref{fig:sensitivity_delay_constraint_bars} analyzes system sensitivity to varying delay constraints from 2.0 to 4.0 seconds. Fig.~\ref{fig:sensitivity_delay_constraint_bars}(b) demonstrates that only centrally trained methods maintain full constraint compliance, with HC-MAPPO-L achieving consistent satisfaction across all thresholds while HC-IPPO-L violates stricter constraints by 12-29\%. Fig.~\ref{fig:sensitivity_delay_constraint_bars}(a) confirms HC-MAPPO-L's cost efficiency advantage, maintaining the lowest user cost between 158.71 and 163.15 while outperforming Heuristic-MAPPO-L by 8.0-12.0 percentage points across the constraint range. Moreover, the proposed algorithm exhibits intelligent resource allocation through adaptive energy-privacy trade-offs. As delay constraints relax from 2.0 to 3.5 seconds, Fig.~\ref{fig:sensitivity_delay_constraint_bars}(c) and (d) show HC-MAPPO-L strategically increases energy consumption by 18\% to enable deeper model splits, effectively reducing delay violation penalties while maintaining cost efficiency. Beyond 3.5 seconds, the constraint becomes non-binding and system performance stabilizes, demonstrating the method's capability to dynamically balance multiple objectives under varying QoS requirements.

\textbf{Impact of Computation Capacity}: Fig.~\ref{fig:sensitivity_user_compute_bars} and~\ref{fig:sensitivity_server_compute_bars} demonstrate the asymmetric adaptation strategy of HC-MAPPO-L under heterogeneous computation capacities. With increasing user computation capacity from 10-20 to 70-80 GFLOPS, HC-MAPPO-L effectively exploits local resources by deepening model partitions, resulting in a 62\% energy increase from 12.2 to 19.7 Joules alongside a 35\% privacy cost reduction from 18.7 to 12.1. This intelligent budget reallocation maintains stable total costs between 160.5-164.3, outperforming competitors that exhibit 2.7-9.0\% cost variations under the same conditions. In contrast, all algorithms show minimal sensitivity to server-side compute capacity variations from 400 to 3000 GFLOPS, with HC-MAPPO-L maintaining near-constant performance with only 1.2\% cost fluctuation. This asymmetric response pattern confirms that server computation does not constitute the primary system bottleneck, and validates our framework's efficient resource utilization by avoiding unnecessary offloading that would degrade privacy without significant delay improvements.

\textbf{Impact of Server Storage Capacity}: Fig.~\ref{fig:sensitivity_server_storage_bars} analyzes system sensitivity to server storage capacity variations. Fig.~\ref{fig:sensitivity_server_storage_bars}(a) demonstrates HC-MAPPO-L's storage efficiency with only 7.8\% cost reduction as capacity increases from 1.8-2.2 GB to 4.8-5.2 GB, achieving near-optimal performance at moderate 3.3-3.7 GB capacity through efficient cache prioritization. In contrast, Greedy exhibits 22.9\% cost improvement under the same conditions, indicating reliance on storage over-provisioning. The proposed algorithm maintains superior deployment robustness across storage constraints, with success rates improving from 94\% to 99\% as capacity grows. This performance consistently exceeds baselines by 5-7\%, demonstrating effective model deployment strategy under limited storage conditions while maintaining balanced performance across all evaluation metrics.

\section{Conclusion}\label{sec:conclusion}
This paper proposes a HC-MAPPO-L algorithm that addresses the complex optimization problem for edge-device collaborative inference. By integrating safety-aware Lagrangian relaxation into MAPPO with a three-layer agent hierarchy, our approach co-optimizes model deployment, user-server association, DNN partitioning, and resource allocation strategy, to minimize the weighted cost under delay and resource constraints. The algorithm combines an auto-regressive policy for combinatorial deployment decisions with an attention-based mechanism for dynamic resource allocation, ensuring both scalability and adaptability. Extensive evaluations demonstrate that HC-MAPPO-L robustly satisfies delay constraints while achieving superior cost-performance trade-offs, maintains exceptional scalability with service success rates above 98.5\%, and learns sophisticated adaptation strategies that efficiently leverage heterogeneous resources. Future work will explore dynamic network topologies and fully local collaborative inference through multi-user cooperation, extend to vehicular networks with high mobility and channel dynamics, enhance DNN partitioning for multi-task models with complex topologies, and incorporate active privacy-preserving mechanisms with advanced multi-objective optimization techniques.
\bibliographystyle{IEEEtran}
\bibliography{refs}

\end{document}